\documentclass[twocolumn, times]{aastex631}

\newcommand*{\msun}{\ensuremath{\rm{M}_{\odot}}\text{ }}
\newcommand*    \kms{{\,\rm km\,s^{-1}}}
\newcommand*    \pc{{\,\mathrm{pc}}}
\newcommand*    \kpc{{\,\mathrm{kpc}}}

\newcommand*    \af{a_{\mathrm f}}
\newcommand*    \ah{a_{\mathrm h}}
\newcommand*    \tk{t_{\mathrm k}}
\newcommand*    \vk{v_{\mathrm k}}
\newcommand*    \ve{v_{\mathrm e}}
\newcommand*    \taf{T_{\mathrm f}}
\newcommand*    \tah{T_{\mathrm h}}

\shorttitle{Core formation in elliptical galaxies}
\shortauthors{Khonji et al.}

\begin{document}

\title[Core formation in elliptical galaxies]{Core formation by binary scouring and gravitational wave recoil in massive elliptical galaxies}

\correspondingauthor{Nader Khonji}
\email{n.khonji@surrey.ac.uk}

\author[0000-0002-6633-2185]{Nader Khonji}
\affiliation{School of Mathematics and Physics, University of Surrey, Guildford GU2 7XH, UK}

\author[0000-0002-9420-2679]{Alessia Gualandris}
\affiliation{School of Mathematics and Physics, University of Surrey, Guildford GU2 7XH, UK}

\author[0000-0002-1164-9302]{Justin I. Read}
\affiliation{School of Mathematics and Physics, University of Surrey, Guildford GU2 7XH, UK}

\author[0000-0001-8669-2316]{Walter Dehnen}
\affiliation{Astronomisches Rechen-Institut, Zentrum f{\"u}r Astronomie der Universit{\"a}t Heidelberg, M{\"o}nchhofstra\ss{}e 12-14, Heidelberg 69120, Germany}



\begin{abstract}
Scouring by supermassive black hole (SMBH) binaries is the most accepted mechanism for the formation of the cores seen in giant elliptical galaxies. However, an additional mechanism is required to explain the largest observed cores. Gravitational wave (GW) recoil is expected to trigger further growth of the core, as subsequent heating from dynamical friction of the merged SMBH removes stars from the central regions. We model core formation in massive elliptical galaxies from both binary scouring and heating by GW recoil and examine their unique signatures. We aim to determine if the nature of cores in 3D space density can be attributed uniquely to either process and if the magnitude of the kick can be inferred. We perform $N$-body simulations of galactic mergers of multicomponent galaxies, based on the observed parameters of four massive elliptical galaxies with cores $>0.5\kpc$. After binary scouring and hardening, the merged SMBH remnant is given a range of GW recoil kicks with $0.5$-$0.9$ of the escape speed of the galaxy. We find that binary scouring alone can form the cores of NGC 1600 and A2147-BCG, which are $< 1.3\kpc$ in size. However, the $> 2\kpc$ cores in NGC 6166 and A2261-BCG require heating from GW recoil kicks of $< 0.5$ of the galaxy escape speed. A unique feature of GW recoil heating is flatter cores in surface brightness, corresponding to truly flat cores in 3D space density. It also preferentially removes stars on low angular momentum orbits from the galactic nucleus.
\end{abstract}



\section{Introduction} \label{sec:intro}

The detection of gravitational waves (GWs) from the merger of a stellar mass black hole binary in 2015 \citep{abbott2016observation} heralded a paradigm shift in observational astronomy. Considerable resources are now focused on the detection of lower frequency GWs from supermassive black hole (SMBH) binaries. SMBHs are likely to reside at the centres of all massive galactic nuclei, with evidence for this ranging from dynamical studies of maser emission \citep[e.g.][]{miyoshi1995evidence, kuo2010megamaser}, to the dynamics of AGN accretion discs \citep{macchetto1997supermassive}, to the direct observation by the Event Horizon Telescope collaboration \citep{EHT2019, EHT2022}. 
 
Cold dark matter (DM) cosmologies such as $\Lambda$CDM predict the hierarchical growth of structure by formation and then mergers of DM halos and galaxies \citep{blumenthal1984formation}. Given the presence of SMBHs, such mergers will inevitably lead to the formation of SMBH binaries (BHBs) \citep{begelman1980massive}. If BHBs form and evolve due to interactions with the background stellar population, they will become strong sources of nanohertz GWs, which are detectable using pulsar timing arrays (PTAs; \citealt[e.g.][]{foster1990constructing}, \citealt{sesana2018testing}). Recently, the European PTA with the Indian PTA \citep{antoniadis2023second}, the North American Nanohertz Observatory for GWs (NANAOgrav) \citep{agazie2023nanograv}, the Chinese PTA \citep{xu2023searching} and the Parkes PTA \citep{reardon2023search}, released analysis of their latest datasets showing exciting evidence for a stochastic nanohertz GW background (GWB).  The background signal is consistent with that of a population of BHBs evolving due to GW inspiral in a stellar/gaseous environment \citep{EPTA2023int}.

Elliptical galaxies contain more than half the total stellar mass of the Universe \citep{bell2003optical, read2005baryonic}. The most massive are found in the centres of galaxy clusters and host the most massive SMBHs. Their inner surface brightness profiles show a range of behaviours, with some rising as steep “cusps” and others asymptoting towards flat inner “cores” \citep{ferrarese1994hubble, lauer1995centers}.  The cored ellipticals are systematically brighter ($M_\mathrm{VT} < -21$) and tend to be anisotropic in velocity, with box-shaped isophotes and low rotation. By contrast, the cusped ellipticals are fainter ($M_\mathrm{VT} > -21$), isotropic in velocity, with disc-shaped isophotes and higher rotation \citep{kormendy1996proposed, kormendy2009structure}. Although this was initially thought to be a dichotomy, there is now known to be some overlap between the two types \citep{rest2001wfpc2}. Most cores are relatively small, from tens to a few hundred parsecs in size \citep{byun1996centers, dullo2014depleted, rusli2013depleted}, but a few are greater than $0.5\kpc$ \citep{dullo2019most}.
    
It had long been postulated that elliptical galaxies are formed by mergers \citep{holmberg1941clustering, toomre1972galactic}. However, initial simulations (without SMBHs) could not replicate the larger cores seen in the most massive galaxies \citep{farouki1983hierarchical}. Indeed, \cite{dehnen2005phase} showed that the steepest cusp is retained in mergers of such systems. The likely solution arose from the theory that active galactic nuclei (AGNs) were the result of accretion by a massive black hole \citep{lynden1969galactic,begelman1978fate}. \cite{begelman1980massive} were the first to argue that mergers would lead to BHBs in AGNs. With the addition of central black holes to the precursor galaxies in their simulations, \cite{ebisuzaki1991merging} found an increase in core size of the remnant.

Observational evidence indicates that quiescent elliptical galaxies were already formed at $z \sim$ 2, but were much more compact than at $z = 0$, with a typical size $\sim 1\kpc$ \citep{van2008confirmation}. Hence, their current size is likely due to major \citep{naab2006properties} and/or minor \citep{naab2009minor} dry mergers. 

In summary, there is a high probability that giant elliptical galaxies are the end product of mergers of galaxies containing SMBHs. \cite{begelman1980massive} first described the likely processes in such mergers which, for gas-poor mergers, can be divided into three stages: (i) dynamical friction against the stellar population and dark matter, which brings the galaxies and SMBHs together, leading to the formation of a BHB; (ii) three-body interactions between stars/dark matter and the binary, often resulting in stellar ejections; and (iii) GW inspiral and coalescence of the binary, with the potential for GW recoil that can push the merged BHB out of the centre of the galaxy and, in extreme cases, even unbind it. 
    
Dynamical friction \citep{chandrasekhar1943dynamical} is dominated by long range (kiloparsec-scale) encounters between each SMBH and the surrounding matter. As a SMBH moves through a field of stars and dark matter, they are pulled to form a `gravitational wake' behind it, slowing the SMBH with the transfer of energy and angular momentum.  In a major merger, this causes the SMBHs to sink into the potential well of the remnant \citep{antonini2011dynamical}. Eventually, they become gravitationally bound and form a binary \citep{valtaoja1989binary}. From around this time, there is rapid hardening of the binary as dynamical friction wanes and short range (parsec-scale) three-body interactions between the SMBH binary and stars on low angular momentum orbits dominate \citep{quinlan1996dynamical, sesana2006interaction}. Simulations of a binary system with low mass intruders show that the vast majority of intruders undergo slingshot ejection from the system, with a concomitant increase in binding energy of the binary \citep{hills1983effect}. Since this process leads to a central mass deficit \citep{merritt2006mass} and reduced central density, it is known as `binary scouring'. It is the most established mechanism for core formation and has been shown to occur in simulations \citep[e.g.][]{quinlan1997dynamical, milosavljevic2001formation, GM2012} and has been shown to preferentially remove stars on low angular momentum orbits \citep{thomas2014dynamical}.
    
Low angular momentum stars and gas populate a region in phase space known as the `loss-cone'. In order for binary hardening to continue past the initial rapid phase and until GW emission becomes significant, the loss-cone must be continually replenished. For an idealised spherically symmetric galaxy, this can only occur by collisional two-body relaxation, a process typically occurring on timescales longer than a Hubble time \citep{makino2004evolution, berczik2005long}. This has led to the `final parsec problem', an envisioned stalling of the hardening of the binary at roughly parsec-scale separations due to the lack of loss-cone refilling in spherical galaxies. However, simulations of galactic mergers from early times show that collisionless loss-cone refilling due to torques and angular momentum diffusion in non-spherical systems lead to efficient hardening of the binary to the GW phase
\citep{khan2011efficient, gualandris2011long, vasiliev2014}. All merger remnants are somewhat triaxial \citep{bortolas2018}, and even a modest triaxiality (e.g. axis ratios 1:0.9:0.8) is sufficient for collisionless loss-cone refilling and coalescence within a Hubble time \citep{vasiliev2015, gualandris2017collisionless}. 

GWs also provide an alternative method for core formation. Asymmetric emission of linear momentum leads to a `recoil kick' on the newly formed SMBH merger remnant. In the non-spinning case, the kick velocity depends solely on the mass ratio:
\begin{equation}
    \label{eq:mass_ratio}
    q = \frac{m_2}{m_1} \leq 1 ,
\end{equation}
where $m_1$ and $m_2$ are the larger and smaller SMBH masses respectively. Numerical simulations by \cite{gonzalez2007maximum} showed the maximum recoil kick in these circumstances is relatively slow at $\sim 175\kms$ for $q \approx 0.36$. However, in the spinning case, the recoil depends on both $q$ and the configuration of the spins. \cite{campanelli2007maximum} found the combination of $q=1$ and spins aligned and anti-aligned with the orbital plane gives a maximum recoil velocity of $\sim 4000 \kms$. In the most extreme cases, where the components of the spin in the orbital plane are of equal magnitude but opposite in sign, and the components out of the plane are equal in magnitude and sign, they can reach $5000\kms$ \citep{lousto2011hangup}, clearly exceeding the escape speed for the galaxy.

It is expected that in most cases kicks will be modest and the newly formed SMBH and core will oscillate back and forth about their common centre of mass. The oscillations are gradually damped as energy is transferred to the stars and dark matter, mostly during passages through the centre, until thermal equilibrium is reached. These interactions result in displacement or even ejection of stars, similar to that in binary scouring, enlarging any pre-existing core formed by the binary and potentially leading to very large cores \citep{gualandris2008ejection}. 

Stalling is another potential mechanism for enlargement of a core \citep{goerdt2010core}. \cite{read2006dynamical} show that dynamical friction would fail at the edge of a constant density core, leading to stalling of an infalling object such as a SMBH. This implies that if a galaxy with a flat core undergoes a further merger, the infalling SMBH may stall and not achieve coalescence. Such `stalled perturbers' could be responsible for the `knots' seen in A2261-BCG \citep{bonfini2016quest, nasim2021formation}. Subsequent infalls into a galaxy with a stalled SMBH could also lead to multiple SMBH systems \citep{lousto2008foundations, liu2011cosmic, kulkarni2012formation}, with the potential for slingshot ejection of a SMBH \citep{iwasawa2006evolution}. 
Hence, the size and flatness of a core has important implications.  Even if the core is not completely flat, the time for the binary to proceed to coalescence could be significantly increased. This is likely to affect the SMBH GWB signal which PTAs are endeavouring to measure \citep{sesana2013insights, ravi2014binary, sampson2015constraining}.
    
For the reasons given above, it is important to be able to accurately determine core size and flatness of the central density profile. This can be achieved by fitting observed luminosity profiles to an appropriate model. There are two models in common use. The Nuker profile \citep{lauer1995centers} uses inner and outer logarithmic slopes separated at the break radius ($r_\mathrm{b}$), which is taken as the size of the core. The inner logarithmic slope ($\gamma$) indicates the flatness of the core. The Nuker profile was designed to fit the central part of the light profile. \cite{graham2003new} showed that the Nuker parameters are dependent on the radial extent of the galaxy used for fitting and may overestimate $r_\mathrm{b}$. They introduced the core-S\'ersic profile, which replaces the outer power law with a S\'ersic function \citep{sersic1963influence}.  

To a large extent, recent simulations support core formation by binary scouring as the main mechanism to explain observed cores in large elliptical galaxies \citep{rantala2018formation, frigo2021two, dosopoulou2021galaxy}. However, it remains unclear whether scouring alone can explain core formation in all cases \citep{nasim2021formation}. \cite{dullo2019most} published core-S\'ersic fits to observations of 12 galaxies with cores larger than $0.5\kpc$, with only two of these having a central slope $\gamma >$ 0.15. Cores such as these are found in the most massive and luminous elliptical galaxies, usually brightest cluster galaxies (BCGs), which contain the most massive SMBHs. Given the processes described above, galaxies with larger SMBHs might be expected to have larger cores.  It is important to test whether binary scouring alone can produce cores of this size and flatness. If not, GW recoil, stalling infallers and multiple SMBH systems are all potential additional mechanisms for the formation of these large, flat cores, as already noted in \citet{nasim2021formation}.

First discovered by \cite{postman2012brightest}, the largest core in \cite{dullo2019most}, and the largest known core to date, is that of A2261-BCG. Using core-S\'ersic fitting, they obtain $r_\mathrm{b} \sim 3\kpc$, with a flat surface brightness profile ($\gamma$ = 0.0). Simulations of core formation based on this galaxy were performed  by \cite{nasim2021formation}, finding that a core of $0.5-1\kpc$ could be explained by binary scouring alone. Although it was previously thought that the mass deficit after multiple mergers is proportional to the number of mergers \citep{merritt2006mass}, i.e. that cores are enlarged by each merger, \cite{nasim2021formation} found that there is minimal increase in both core size and mass deficit with subsequent major or minor mergers. It becomes progressively harder to carve a core once one is already present and the central density has been lowered.
This implies that an additional mechanism is required to explain the largest observed cores. They explored GW recoil following binary coalescence as the most likely mechanism, and showed that it can be very efficient at enlarging a pre-existing core.
Furthermore, \cite{nasim2021formation} showed that cores retain shallow cusps after scouring, especially spatial density, even after sequential dry mergers, and GW recoil is required to form truly flat cores. 

In this paper we study the physical processes involved in core formation in a range of galaxies, to understand how large cores are formed and what density profiles they produce, i.e. shallow cusps or flat profiles. We test whether binary scouring and GW recoil have unique observational signatures that can predict if GW recoil has occurred in a given galaxy. If recoil is required, the signatures could also provide an indication of the strength of the required kick. To this end, we simulate major mergers with parameters based on four galaxies from \cite{dullo2019most}, including A2261-BCG, with a range of core sizes above $0.5\kpc$. We investigate whether and how much GW recoil is required to achieve the observed core sizes, and if GW recoil produces fully flat cores in spatial density. We compare our results to observations of a sample of cored galaxies. For the first time, we attempt to explain differences in the phase space density of stars as a result of dynamical friction/stellar hardening and GW recoil by analysing the energy and angular momentum transfer in the two processes. We find that, similarly to \cite{nasim2021formation}, the two galaxies with cores greater than $2\kpc$ require GW recoil to achieve their observed size. Furthermore, recoil produces flatter cores than binary scouring alone, and needs to be invoked to explain the observed flatness in the profiles of all the four galaxies. We find that the required GW kicks are modest at less than half the escape velocity of the galaxies. Finally, we show that both binary scouring and recoil preferentially remove low energy stars, but that GW recoil in particular ejects low angular momentum stars from the core. This is why GW recoil produces truly flat, constant density, cores in 3D. Core formation by the stalled perturber and multiple SMBH scenarios are not considered further here, but may be revisited in future work.

This paper is organised as follows. The selection of galaxies, merger and GW recoil simulation methodology and density profile fitting procedure are described in Section \ref{sec:methods}. The formation and evolution of BHBs, core-fitting and study of the energy and angular momentum exchanges are presented in Section \ref{sec:results}. Finally, discussion and conclusions of the comparison between binary scouring and GW recoil are presented in Section 4.

\section{Methods} \label{sec:methods}
We performed equal-mass simulations of galactic mergers to study the formation of large cores in giant elliptical galaxies and to compare the processes of binary scouring and GW recoil. Four galaxies with observed large cores were selected for modelling. These were chosen to have a range of core sizes ($\sim 0.5 -3\kpc$) and SMBH masses, as well as variation in other parameters such as effective radius, S\'ersic index and bulge mass ($M_*$). Observational data was taken from \cite{dullo2019most}, who performed core-S\'ersic fits to their surface brightness profiles from both Hubble Space Telescope and ground-based images. SMBH mass estimations were available from their $r_\mathrm{b}$-$M_\bullet$ relation, with the exception of NGC 1600, which has a direct measurement. \cite{dullo2019most} also provided $M_*$ estimations using the $M_*$-luminosity relation from \cite{worthey1994comprehensive}. Although there is significant scatter in core-SMBH relations, \cite{dullo2019most} found that the $r_\mathrm{b}$-$M_\bullet$ relation was the most reliable for the most massive galaxies, compared to luminosity and velocity dispersion. The galaxy parameters are summarised in Table \ref{tab:observed}.

\begin{deluxetable*}{lcccccccc}
\tablecaption{Parameters of the selected galaxies.}
\label{tab:observed}           
\tablehead{
\colhead{Galaxy} & \colhead{Type} & \colhead{$r_\mathrm{b}$} & \colhead{$\gamma$} & \colhead{$\alpha$} & \colhead{$n$}  & \colhead{$r_\mathrm{e}$} & \colhead{$M_\bullet$}\hspace{-0.2cm}\tablenotemark{\scriptsize{a}} & \colhead{$M_*$}\hspace{-0.2cm}\tablenotemark{\scriptsize{b}}\\ 
&           & \colhead{[kpc]}    &            &           &       & \colhead{[kpc]} & \colhead{[$10^{10} M_\odot$]}   & \colhead{[$10^{12} M_\odot$]}  }
\startdata
NGC 1600    & Isolated  & 0.65     & 0.04       & 2         & 6.3   & 22.8  & 1.70                &   1.51               \\
A2147-BCG   & BCG       & 1.28     & 0.14       & 2         & 6.4   & 31.8  & 2.63                &   1.34               \\
NGC 6166    & BCG       & 2.11     & 0.14       & 2         & 9.0   & 83.1  & 4.79                &   3.39               \\
A2261-BCG   & BCG       & 2.71     & 0.00       & 5         & 2.1   & 17.6  & 6.45                &   4.07               \\
\enddata
\tablerefs{\cite{dullo2019most}}
\tablecomments{First five parameters are from core-S\'ersic fits.}\vspace{-0.2cm}
\tablenotetext{a}{Black hole masses from the $M_\bullet-r_\mathrm{b}$ relation except NGC 1600 (directly measured)}\vspace{-0.2cm}
\tablenotetext{b}{Bulge mass from spheroid luminosity.}
\end{deluxetable*}

\begin{deluxetable*}{lccccccccc}
\tablecaption{Parameters of precursor galaxies.}             
\label{tab:precursors}              
\tablehead{
\colhead{Precursor} & \colhead{$M_\bullet$} & \colhead{$M_*$} & \colhead{$M_{200}$} & \colhead{$n$} & \colhead{$r_\mathrm{e}$} & \colhead{$N_*$} & \colhead{$N_\mathrm{DM}$} & \colhead{$N$} & \colhead{Maximum Stellar Resolution}\\
                        & \colhead{[$10^{10} M_\odot$]} & \colhead{[$10^{12} M_\odot$]} & \colhead{[$10^{14} M_\odot$]} &  & \colhead{[kpc]} & & & & \colhead{[$10^6 M_\odot$]} }
\startdata
NGC 1600 P   & 0.85                  & 0.76                  & 0.79                  & 6.3   & 11.4               & \phantom{5}94902  &   959532  & 1054434   & 0.84\\
A2147-BCG P  & 1.32                  & 0.67                  & 1.00                  & 6.4   & 15.9               & \phantom{5}68065  &   986478  & 1054543   & 1.04\\
NGC 6166 P   & 2.40                  & 1.69                  & 1.00                  & 9.0   & 41.7               & 153422            &   901543  & 1054965   & 1.17\\
A2261-BCG P  & 3.23                  & 2.04                  & 1.00                  & 2.1   & \phantom{5}8.8     & 176091            &   830565  & 1054436   & 1.20\\
\enddata
\end{deluxetable*}

\vspace{-1.7cm}
Multicomponent equilibrium models of precursor galaxies were realised using the AGAMA action-based modelling library \citep{vasiliev2019agama}. Potential models for all three components (SMBH, stellar bulge and DM halo)  and density models for the bulge and halo were made. 
For the stellar bulge, a S\'ersic  profile \citep{sersic1963influence} was used, which can be represented as:
\begin{equation}
\label{eq:sersic}
I(r) = I(0) \, e^{-b_n(r/r_\mathrm{e})^{1/n}} \ .
\end{equation}
This has three free parameters:  the central surface brightness $I(0)$, the effective radius $r_\mathrm{e}$, and the S\'ersic index $n$. Here, $b_n$ is a function of $n$ that ensures $r_\mathrm{e}$ is the half-light radius ($b_n \approx 2n - 1/3$). AGAMA uses a deprojected profile in which the bulge mass can be used as a proxy for luminosity.

For the DM halo, we used the Navarro-Frenk-White profile \citep{navarro1997universal}:
\begin{equation}
\label{eq:nfw}
\rho(r) = \frac{\rho_\mathrm{s}}{(r/r_\mathrm{s}) \,(1 + r/r_\mathrm{s})^2} \ ,
\end{equation}
where $\rho_\mathrm{s}$ is the characteristic density and $r_\mathrm{s}$ is the scale radius. The latter is defined by $r_\mathrm{s}$=$r_{200}/c$, where $r_{200}$ is the radius at which the mean enclosed density is $200$ times the critical density, and $c$ is the halo concentration. The concentration parameter $c$ is calculated using the concentration-mass relation at redshift $z = 0$ from \cite{dutton2014cold}:
\begin{equation}
\label{eq:concentration}
\mathrm{log} \, c = 0.905 - 0.101 \, \mathrm{log} \, \bigg(\frac{M_{200}\, h}{10^{12} \, M_\odot}\bigg) \ ,
\end{equation} 
where $h$ has the usual relation to the Hubble constant ($H_0 =$ 100 $h\, $km s$^{-1}$ Mpc$^{-1}$). $M_{200}$ is the enclosed mass at $r_{200}$ given by:
\begin{equation}
\label{eq:m200}
M_{200} = 200 \,  \rho_{\mathrm{crit}} \, \case{4}{3}\pi \, R_{200}^3 
\end{equation}  
where $\rho_0$ is a characteristic density which is dependent only on $c$ and $\rho_{\mathrm{crit}}$. Clearly, once $M_{200}$ is chosen, all other halo parameters can be calculated.

Spherical distribution functions were created using the potential and density models. These were then sampled to provide ergodic $N$-body galaxy models.

We simulated the formation of each chosen galaxy by a major merger of two precursor galaxies, identified by the name of the observed galaxy with the suffix `P', with parameters (Table \ref{tab:precursors}), based on the observed values (Table \ref{tab:observed}). The precursor SMBH and stellar bulge masses were assumed to be half that of those of the merger remnant. The value of $r_\mathrm{e}$ used for the bulge was half that of the core-S\'ersic fit by \cite{dullo2019most}, since the virial theorem predicts $r_\mathrm{e}$ should double in an equal-mass dissipationless merger \citep{naab2009minor}. Since three of the galaxies are BCGs, their halos are indistinct from the neighbouring galaxies in their clusters. Hence $M_{200}$ was estimated at $\sim 10^{14} M_\odot$ using the $M_* - M_{200}$ relation from \cite{correa2020dependence}. NGC 1600 is unusual in that it is an isolated galaxy, and hence an observed value $M_{200}=1.58$x$10^{14} M_\odot$ is available \citep{goulding2016massive}, of which half was used for the precursors. 

The total particle number was $N\sim 10^6$ for each precursor, to optimally balance overall resolution and computational resources. The numbers of bulge and halo particles were then chosen to give a halo-to-bulge particle mass ratio (PMR) of $\sim$10. This was to balance the need for sufficient resolution of the core with that to avoid significant dynamical segregation of the halo particles, which may result from a PMR that is too high. Core resolution was further increased by the use of the `mass-refinement' scheme of \citet{attard2024}. This works by oversampling the particle distribution by a factor of ten and dividing each particle type into four concentric zones. The inner zone containing $1\%$ of particles is left untouched. The outer zones containing $11.5\%$, $38.5\%$ and $49\%$ of particles are increased in particle mass by factors of $\sim 2.53$, $10$ and $40$ respectively. Then the particle numbers are correspondingly reduced to restore $N$ to its initial value and to ensure that the density profile is unaffected. The technique allows to increase central resolution at the same overall $N$, and has been shown to reduce stochastic effects in simulations of galactic mergers \citep{attard2024}. The maximum (central) mass resolution for each galaxy is shown in Table \ref{tab:precursors}.

\begin{deluxetable}{lccc}
\tablewidth{\columnwidth}
\tablecaption{Orbital parameters at the end of scouring.}
\label{tab:orbital}
\tablehead{
\colhead{Remnant} & \colhead{\hspace{.6cm}$t_\mathrm{k}$} & \colhead{\hspace{.6cm}$\ah$} & \colhead{\hspace{.6cm}$e$} \\
& \colhead{\hspace{.6cm}[Myrs]} & \colhead{\hspace{.6cm}[pc]} & 
}
\startdata
NGC 1600 M  & \hspace{.6cm}310   & \hspace{.6cm}23    & \hspace{.6cm}0.18  \\
A2147-BCG M & \hspace{.6cm}385   & \hspace{.6cm}38    & \hspace{.6cm}0.09  \\  
NGC 6166 M  & \hspace{.6cm}285   & \hspace{.6cm}30    & \hspace{.6cm}0.27  \\
A2261-BCG M & \hspace{.6cm}470   & \hspace{.6cm}30    & \hspace{.6cm}0.21  \\
\enddata
\end{deluxetable}

For the mergers, denoted by the name of the galaxy being modelled with the suffix `M', the precursor galaxies were placed on a highly elliptical orbit ($e = 0.95$). This was to replicate the almost radial mergers seen in cosmological simulations \cite[e.g.][]{khochfar2006orbital}.
The initial separation between the two galaxies was set to $d=80\kpc$ to ensure that stellar bulges are still well separated.

All $N$-body simulations were performed with GRIFFIN \citep{dehnen2014fast}, a fast multipole method $N$-body code optmised to achieve force errors comparable to direct summation while requiring only $\mathcal{O}(N)$ operations. All interactions were softened with a Plummer-type softening parameter of $3\pc$ for any interaction involving the SMBHs and $30\pc$ for all other bulge and halo particle interactions. 

The SMBHs were followed through the galactic merger and the phase of dynamical friction to formation of a binary and further hardening by encounters with stars. Once the binary had hardened, the SMBHs were assumed to have merged due to GW emission, and the SMBH remnant was placed at the centre of mass of the binary prior to coalescence. They were then given a `GW kick', arbitrarily in the $x$-direction, of $0.5$, $0.7$ or $0.9$ as a proportion of the escape speed $v_\mathrm{e}$ from that galaxy. This was calculated numerically as $v_\mathrm{e} = \sqrt{-2\Phi}$, where $\Phi$ is the total potential of the bulge and halo. The simulation was then continued until the oscillatory motion of the SMBH remnant had settled to the level of Brownian motion.  The `Brownian velocity' $v_{\mathrm{B}}$ can be determined simply from equipartition of energy and can be written as 
\citep{smoluchowski1906essai, merritt2001brownian, bortolas2018}
\begin{equation}
\label{eq:brownian}
v_{\mathrm{B}}^2 = \frac{m}{M_\bullet} \sigma_*^2 \ ,
\end{equation}  
where $m$ is the mass of a stellar particle and $\sigma_*$ is the stellar velocity dispersion.

\section{Results} \label{sec:results}

\begin{figure*}
    \includegraphics[width = \textwidth]{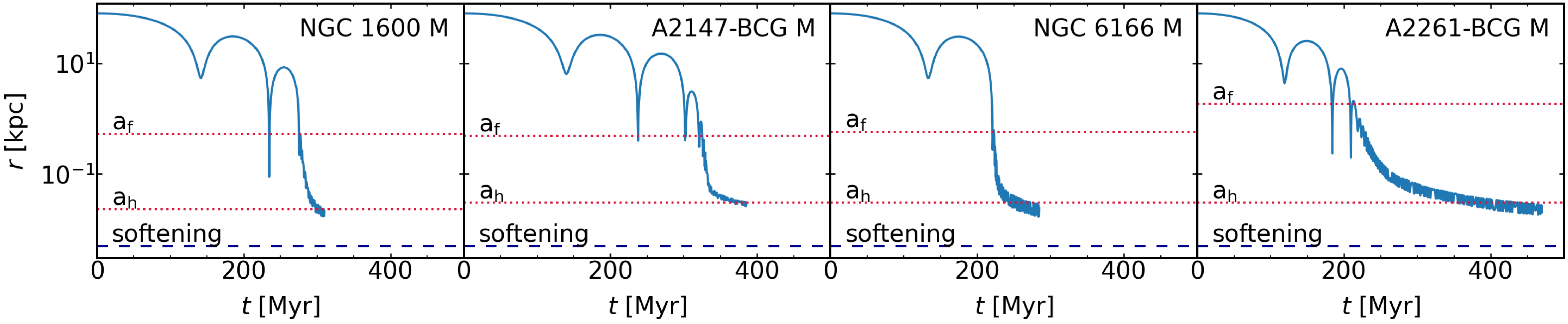}
    \caption{Separation between the SMBHs as a function of time in models NGC 1600 M, A2147-BCG M, NGC 6166 M and A2261-BCG M, showing the three characteristic phases of binary evolution. The dotted lines indicate the critical separations corresponding to $a_\mathrm{f}$, the influence radius of each SMBH, and $a_\mathrm{h}$, the hard binary separation (see equations \ref{eq:af} and \ref{eq:ah}) ; the dashed line indicates the SMBH softening length.}
    \label{fig:distance}
\end{figure*}

\begin{figure*}
    \includegraphics[width = \textwidth]{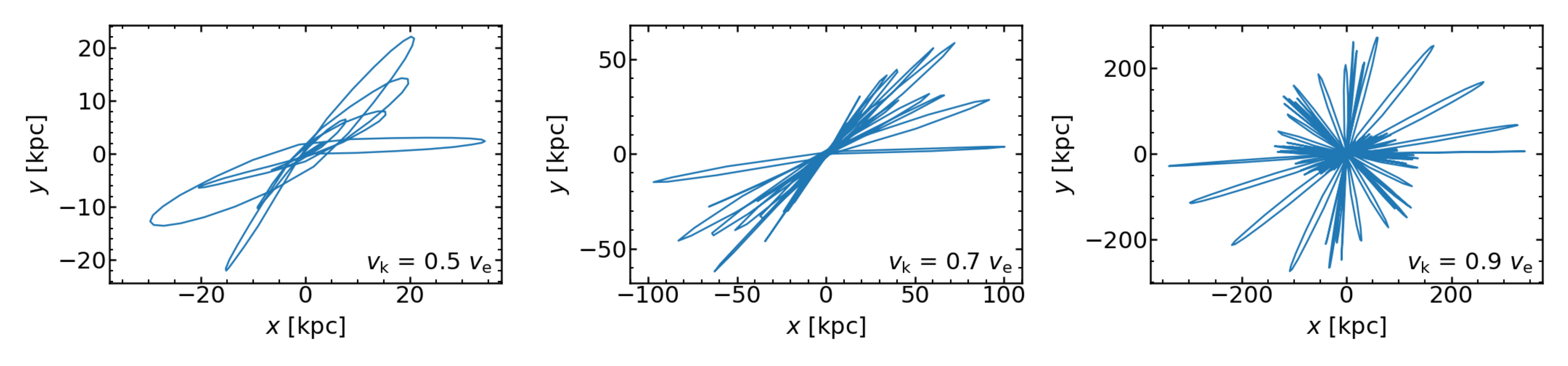}
    \includegraphics[width = \textwidth]{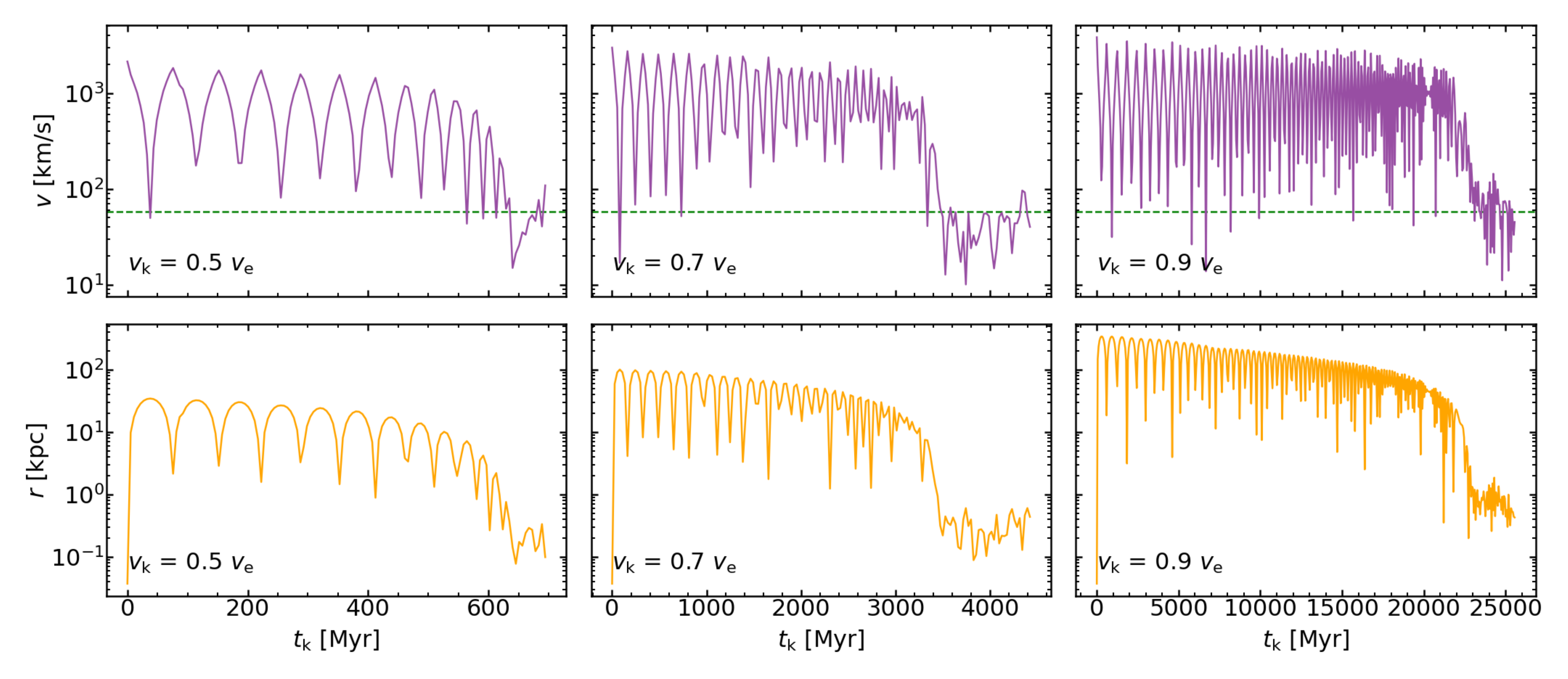}
    \caption{Trajectory (top row), speed (middle row) and 
    distance from the COM of the SMBH remnant (recentred on the COM) in NGC 1600 M simulations from the time the GW kick recoil is applied. Columns (from left to right) show results for $v_\mathrm{k} / v_\mathrm{e}$ of 0.5, 0.7 and 0.9 respectively. The time $t_\mathrm{k}$ is time since the GW kick. The green dashed line indicates the estimated Brownian velocity for each SMBH, computed according to Eq.\,\ref{eq:brownian}. Once the SMBH reaches this velocity, we can consider it settled and in thermal equilibrium in the core of the galactic remnant.}
    \label{fig:trajectory}
\end{figure*}

\begin{figure*}[htb!]
    \includegraphics[width = \textwidth]{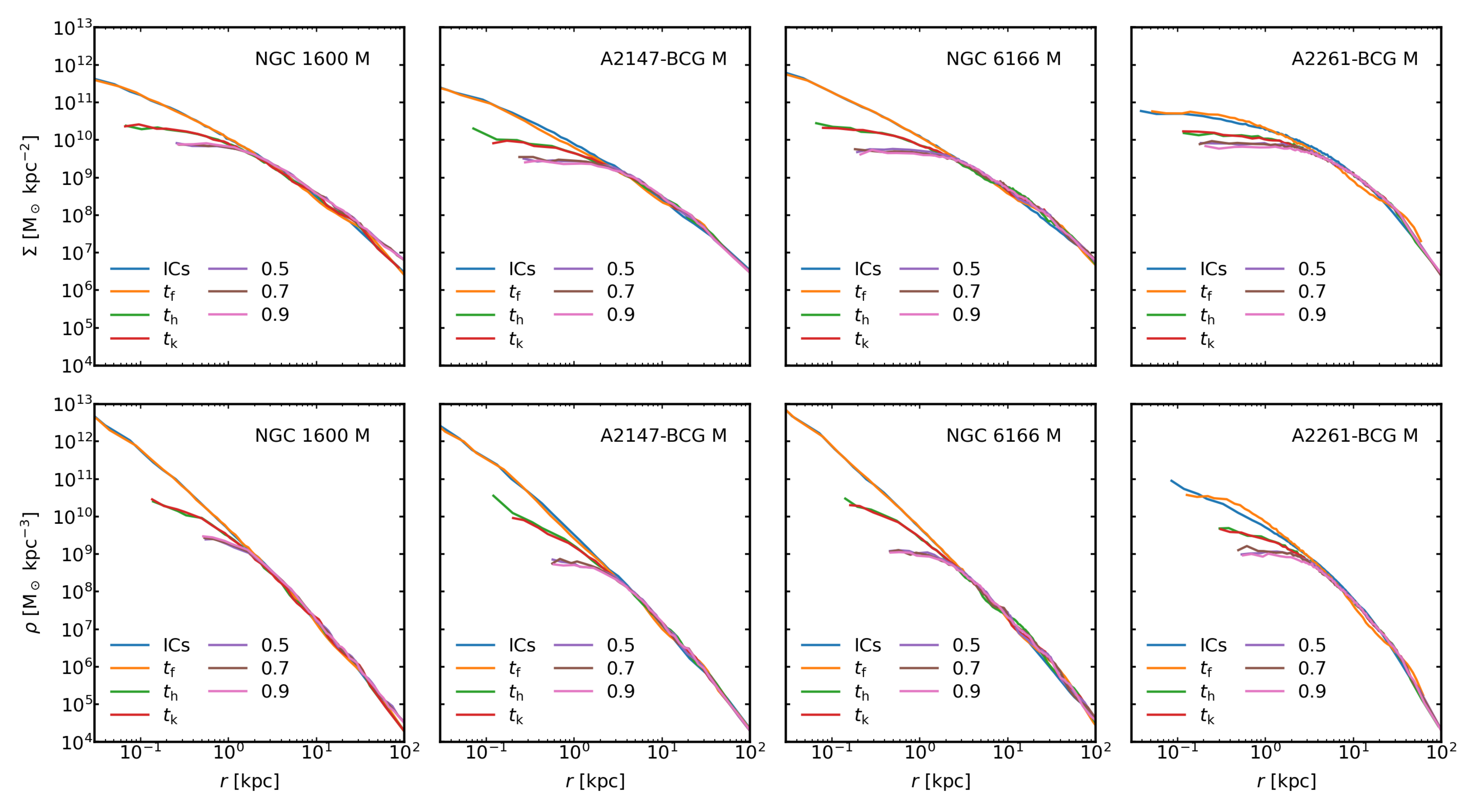}
    \caption{Surface density (top row) and volume density (bottom row) profiles of the stellar component of the four galaxies at different times during the evolution: at the start of the simulations (marked as ICs); at the time $\taf$ when separation $\af$ is reached; at the time $\tah$ when separation $\ah$ is reached; at the time $t_k$ when the GW recoil kick is applied; and at the time the first apocentre after the kick is reached, for each kick velocity (labelled by its value of $v_\mathrm{k}/v_\mathrm{e}$).}
    \label{fig:density_merger}
\end{figure*}

\begin{figure*}
    \includegraphics[width = \textwidth]{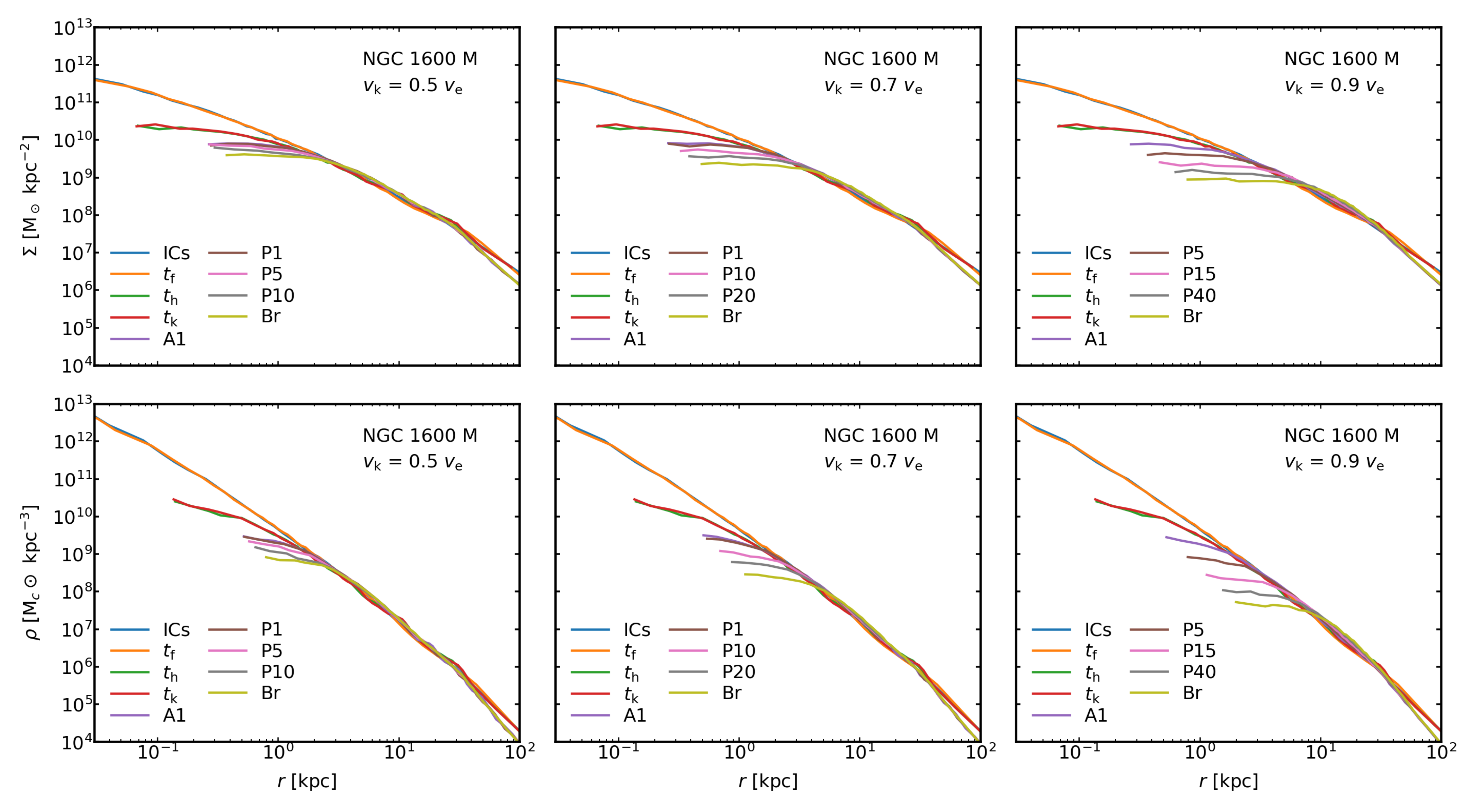}
     \caption{Surface density (top row) and volume density (bottom row) profiles of the stellar component during the merger (from the start of the simulation to GW recoil) and after the GW kick (here `A' and `P' indicate  apocentre and pericentre passages, followed by the number of passages; `Br' indicates the SMBH remnant has settled into Brownian motion) for galaxy NGC 1600 with $v_\mathrm{k}/v_\mathrm{e}=0.5, 0.7, 0.9$, from left to right.}
    \label{fig:density_kicks_NGC1600}
\end{figure*}

\begin{figure*}
    \includegraphics[width = \textwidth]{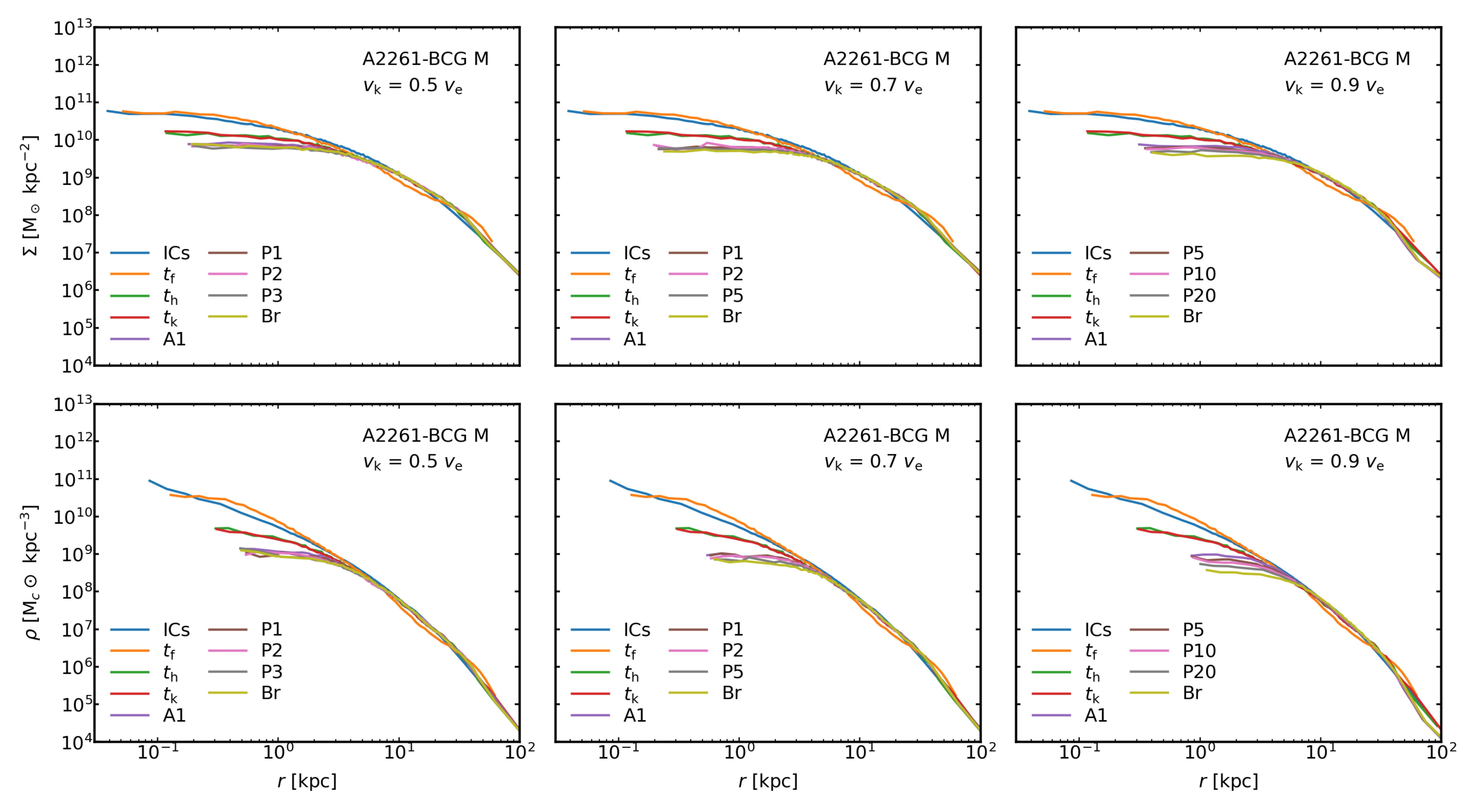}
     \caption{As Fig. \ref{fig:density_kicks_NGC1600}, but for galaxy A2261-BCG M.}
    \label{fig:density_kicks_A2261}
\end{figure*}

\begin{figure}
    \includegraphics[width = \columnwidth]{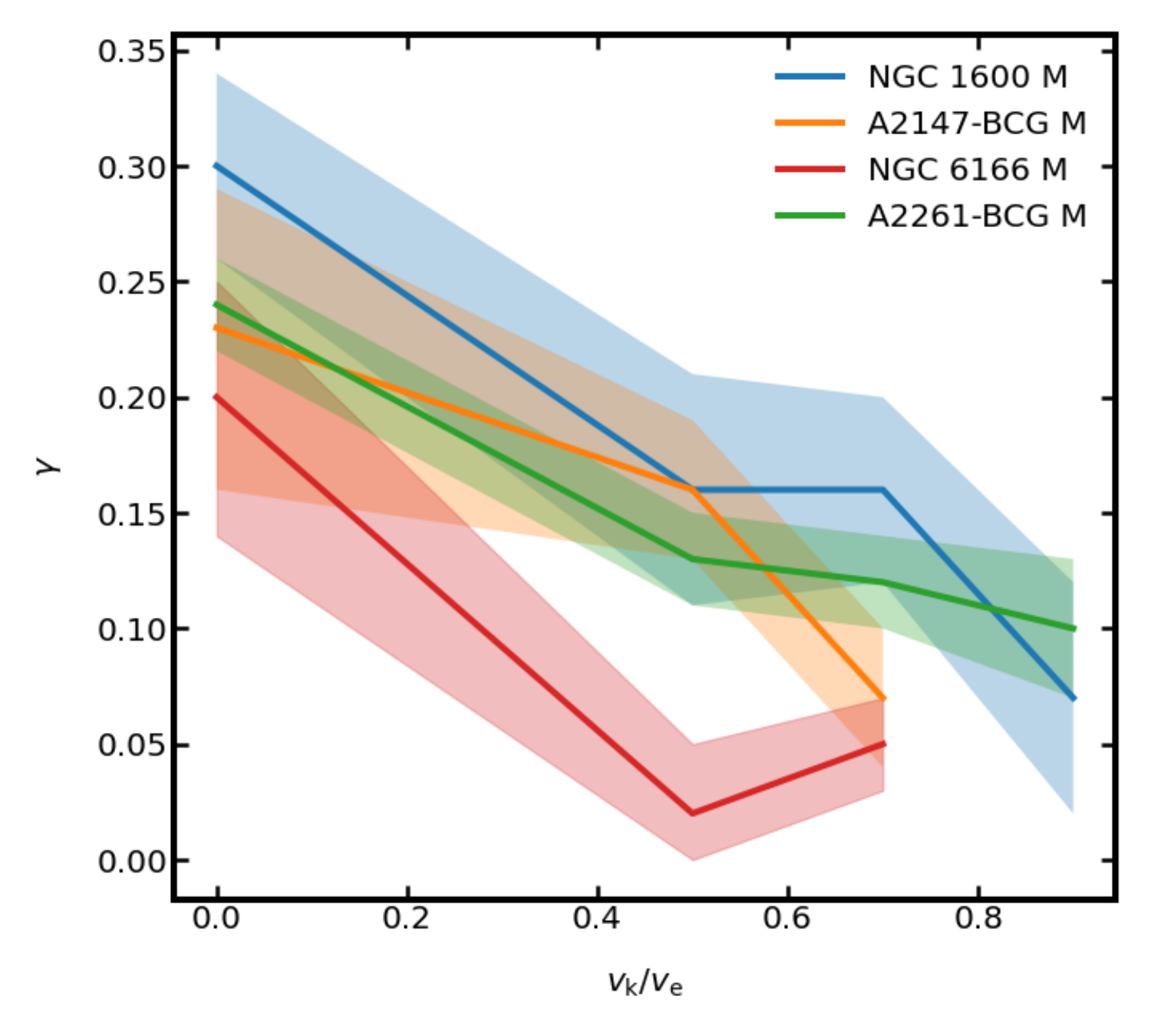}
     \caption{Inner logarithmic slope ($\gamma$) as a function of the ratio of kick velocity ($v_\mathrm{k}$) to escape velocity ($v_\mathrm{e}$) for all four galaxy remnants. A $\vk/\ve$ of zero indicates scouring only. The shaded bands indicate 68\% confidence intervals.}
    \label{fig:gamma_vk}
\end{figure}

\begin{figure*}
\centering
\includegraphics[width = 0.94\textwidth]{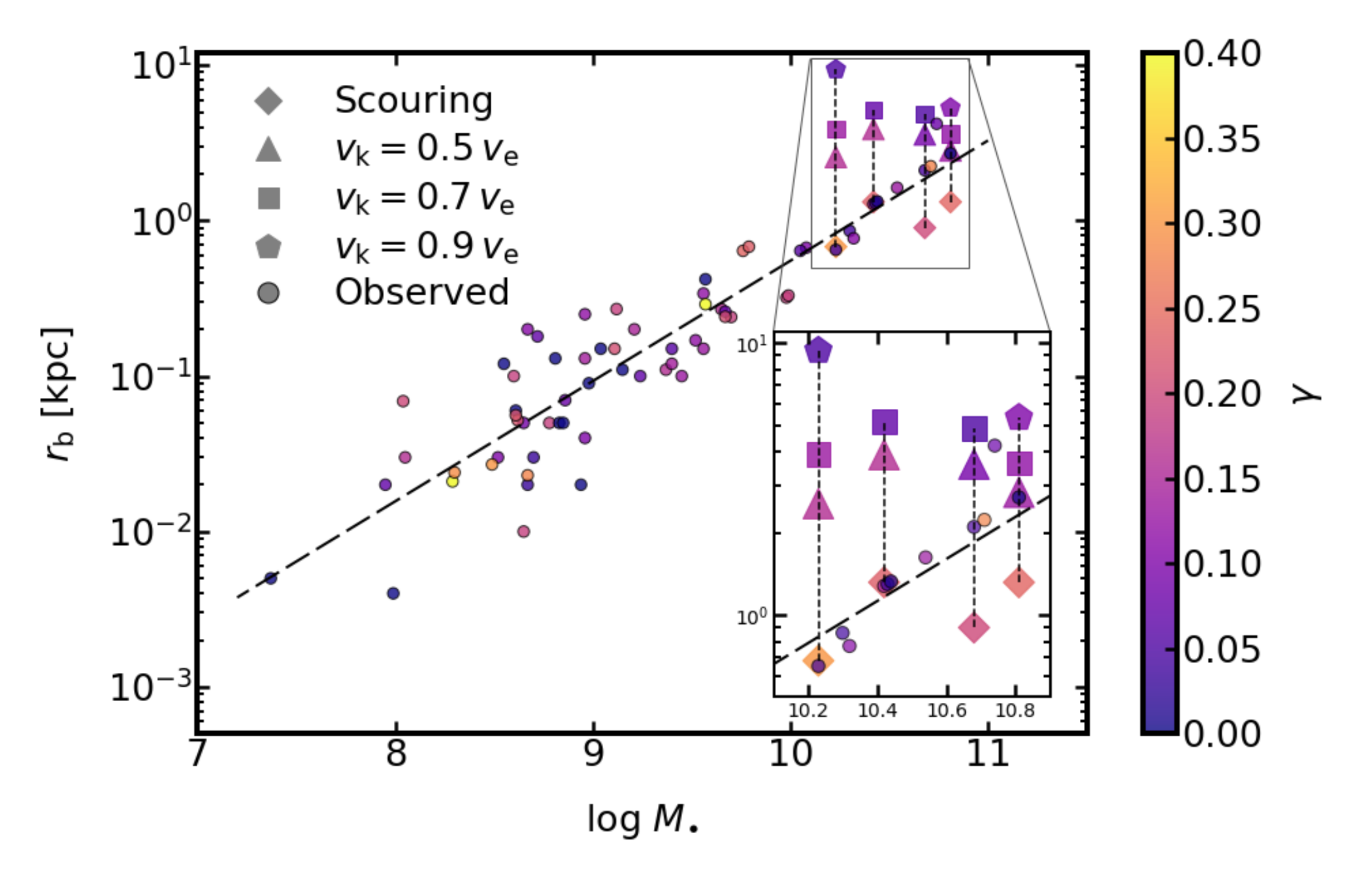}
\caption{Core size and black hole mass scaling relation. Simulation results after scouring are plotted as diamonds and after GW kicks as triangles, squares and pentagons indicating a kick magnitude of 0.5, 0.7 and 0.9 of $v_{\rm k}/v_{\rm e}$ respectively. The scouring and kick results for each merger remnant are joined by a dashed line. The remnants are (from left to right) NGC 1600 M, A2147-BCG M, NGC 6166 M and A2261-BCG M.  The circles are observational fits from \protect\cite{dullo2019most}, \protect\cite{dullo2014depleted} and \protect\cite{rusli2013depleted}.}
\label{fig:rb_mbh}
\end{figure*}

\begin{figure*}
\centering
\includegraphics[width = 0.9\textwidth]{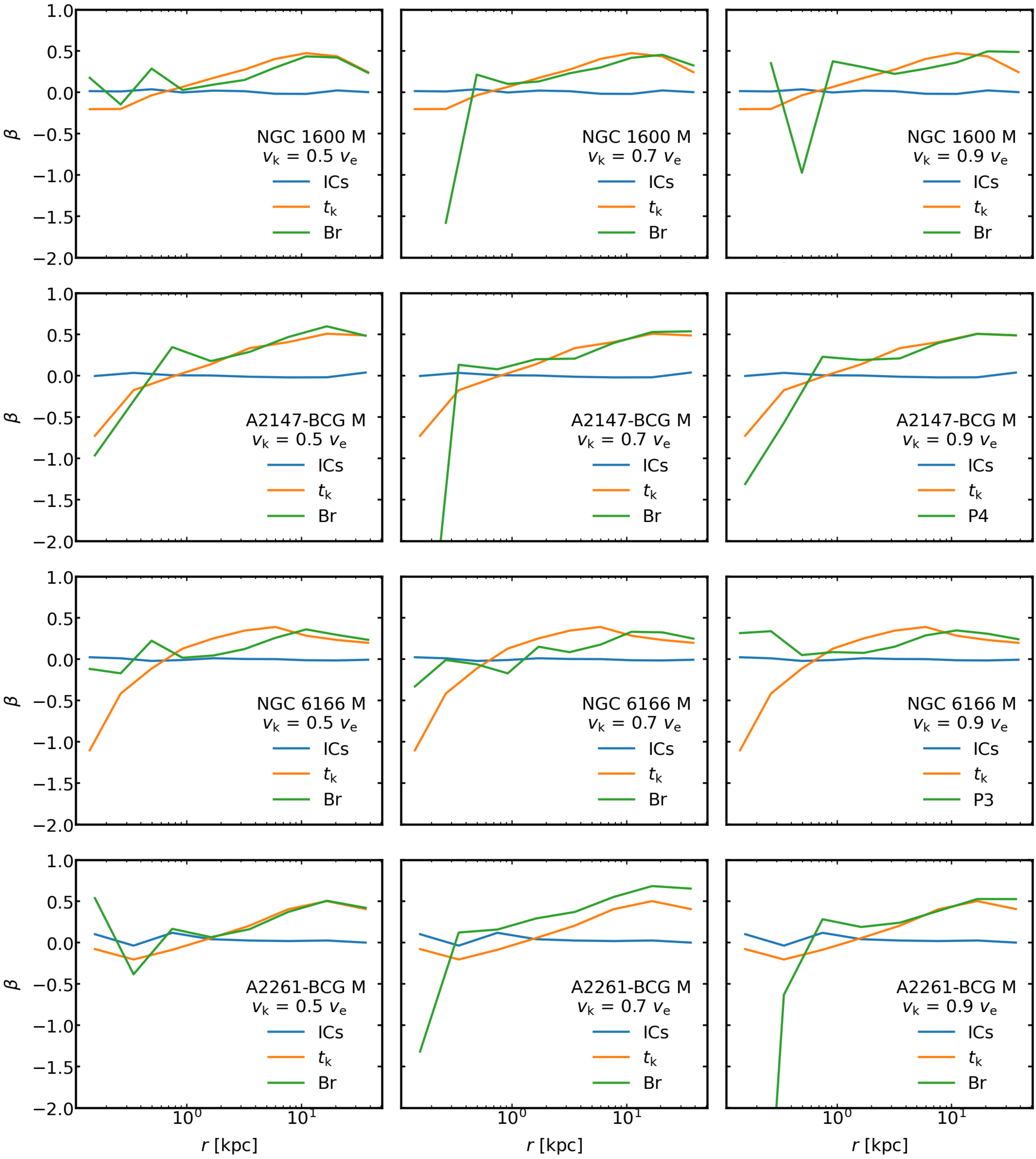}
\caption{Velocity anisotropy parameter ($\beta$) as a function of radius. The rows (from top to bottom) correspond to NGC 1600 M, A2147-BCG M, NGC 6166 M and A2261-BCG M, and the columns (from left to right) to kick magnitudes of 0.5, 0.7 and 0.9 $v_{\rm k}/v_{\rm e}$ respectively. $\beta$ is shown at the start of the merger (`ICs'), at the end of scouring ($t_\mathrm{k}$) and the end of recoil, where the SMBH has reached Brownian motion (Br) or the most recent pericentre if the simulation is unfinished. For the latter, 'P' is followed by the number of passages.}
\label{fig:beta}
\end{figure*}

\begin{figure*}[t!]
    \centering
    \includegraphics[width = 0.9\textwidth]{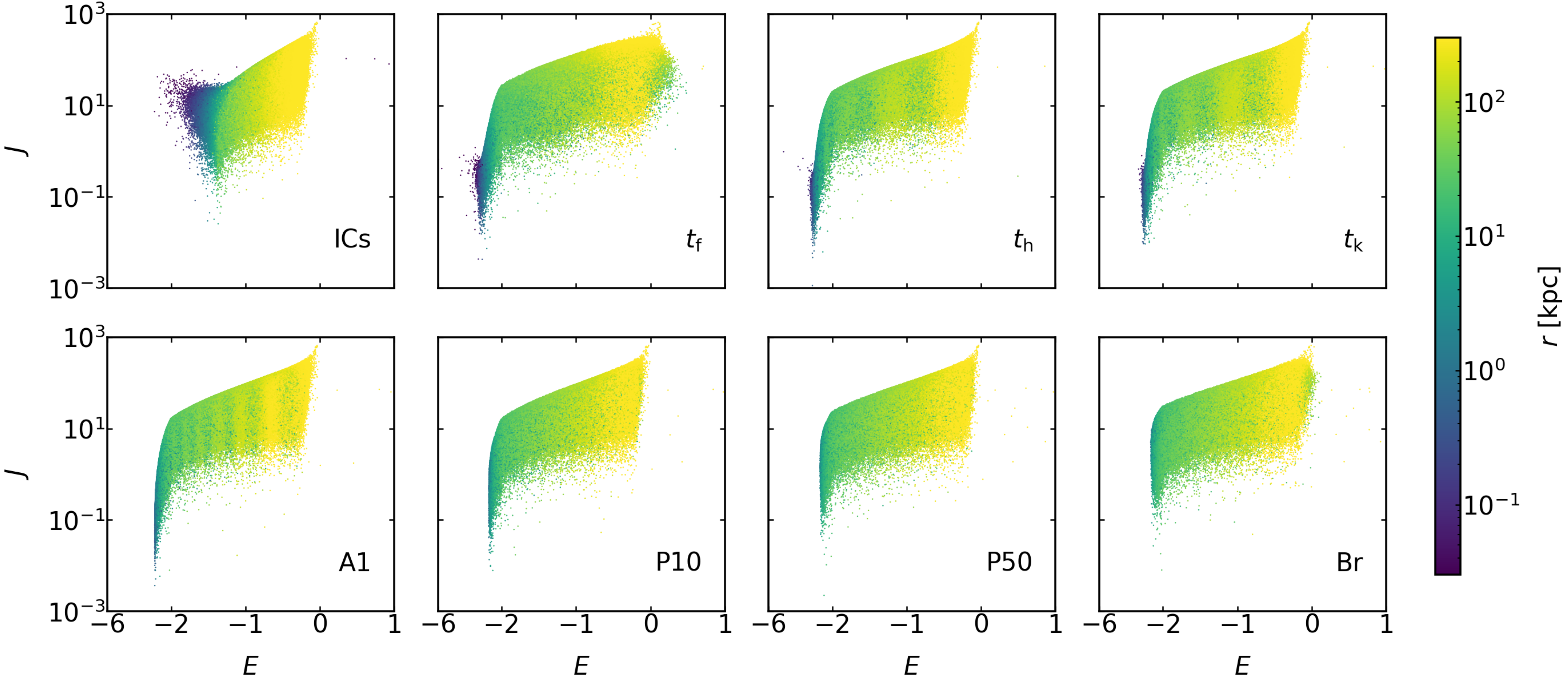}
    \caption{Energy - angular momentum maps for NGC 1600 M during: Initial merger and binary scouring phase (top row); after application of a GW kick velocity $v_\mathrm{k} = 0.9\,v_\mathrm{e}$ (bottom row). `ICs' are the initial conditions at the start of the merger, $t_\mathrm{f}$ and $t_\mathrm{h}$ are the times when the SMBH separation reaches $a_\mathrm{f}$, the influence radius of each SMBH, and $a_\mathrm{h}$, the hard binary separation, respectively (see equations \ref{eq:af} and \ref{eq:ah}), and $t_\mathrm{k}$ is the time when the $N$-body simulation is paused for the SMBH merger and GW recoil; `A' and `P' indicate  apocentre and pericentre respectively, and are followed by the number of passages. `Br' indicates the SMBH remnant has settled into Brownian motion. The colourmap shows the radial distance from the centre of mass of the merger remnant.}
    \label{fig:ej_scourkick}
\end{figure*}

The formation and evolution of the BHB during the mergers is shown in Fig. \ref{fig:distance}, with $t = 0$ at the start of the merger simulation. The first phase, dominated by dynamical friction, ends when the separation reaches $a_\mathrm{f}$, the influence radius of each SMBH (or the secondary SMBH in the case of unequal-mass mergers). It is defined as the separation at which the enclosed stellar mass is twice the (secondary) SMBH mass:
    \begin{equation}
    \label{eq:af}
    M_* \,(r < a_\mathrm{f}) = 2m_2 \,.
    \end{equation}
  A2261-BCG M has a significantly larger $\af$ at $1.85\kpc$, compared to $0.58 \kpc$ for NGC 6166 M. This is likely due to the combination of a higher $M_\bullet$ and a flatter inner S\'ersic profile in A2261-BCG M. 
  
  After this, there is rapid hardening of the binary due to three-body encounters with stars and binary scouring. The specific binding energy exceeds the specific kinetic energy \citep{milosavljevic2001formation} at the hard binary separation:
    \begin{equation}
    \label{eq:ah}
    a_\mathrm{h} = \frac{GM_\bullet}{4\sigma^2} \ ,
    \end{equation}
    where $\sigma$ is the stellar velocity dispersion. In the case of an unequal-mass merger, the reduced mass of the binary would be used. The simulations were paused once they had reached this point for the GW kick to be applied. The time at which the SMBH binary was merged ($\tk$), the value of $\ah$ and the eccentricity ($e$) of the binary at the time of merger are shown for each remnant in Table \ref{tab:orbital}. We note that the circularisation of the binary is consistent with the findings of \cite{fastidio2024eccentricity} for mergers with initial high eccentricity.

The top row of Fig. \ref{fig:trajectory} shows the trajectory of the NGC 1600 M SMBH remnant after the GW kick, relative to the centre of mass (COM) of stars and DM. The middle and bottom rows show the speed of the remnant and distance from the COM with respect to the time of the GW kick at $t = t_\mathrm{k}$. The columns show the results for kicks of $v_\mathrm{k} / v_\mathrm{e}$ = 0.5, 0.7 and 0.9, respectively. Clearly, as $v_\mathrm{k} / v_\mathrm{e}$ increases, the distance to the first apocentre, the number of passages through the centre of the core, and the time taken for the remnant to settle into Brownian motion, increase dramatically. For $v_\mathrm{k} / v_\mathrm{e}$ = 0.9, the remnant makes 103 passages and takes more than a Hubble time to settle. We note that the age of NGC 1600 has been estimated at $7.3 \pm 1.5$ Gyr \citep{terlevich2002catalogue}, which would exclude this largest kick. However, the findings remain relevant for other galaxies with similar parameters.

The surface and volume density profiles of all four galaxies up to the first apocentre of the GW kicks are shown in Fig. \ref{fig:density_merger}. The profiles show no significant change from the initial conditions (`ICs') of the merger until $t_\mathrm{f}$, the time at which the SMBH binary separation reaches $a_\mathrm{f}$. However, as the binary hardens between $t_\mathrm{f}$ and $t_\mathrm{h}$, the time the separation reaches  $\ah$, core formation due to binary scouring is evident as a flattening of the central profiles. Although the cores appear flatter in the surface density (2D) profiles than in the 3D density profiles, they still retain some cuspiness after scouring alone. There is, on the other hand, rapid flattening of the profiles after the GW kicks, largely complete by the time the SMBH remnant reaches its first apocentre. The surface density profiles already appear flat in 2D, corresponding to shallow cusps in 3D. With further passages, the central density gradually reduces whilst the degree of flattening is similar or mildly increased.
We observe no significant difference in the profiles at these times between the different kick values and the cores do not appear to be larger than they were after scouring.

Fig. \ref{fig:density_kicks_NGC1600} shows the profiles for NGC 1600 M during binary scouring and for the entirety of each GW kick. As the kick velocity increases and passages through the core become more frequent, the profiles indicate progressive enlargement of the core. For the largest kick ($v_\mathrm{k}/v_\mathrm{e}$ = 0.9) the core continues to grow even after 50 passages and becomes extremely large. The corresponding profiles for A2261-BCG M are shown in Fig. \ref{fig:density_kicks_A2261}. Although increasing $v_\mathrm{k}/v_\mathrm{e}$ leads to a larger final core size, the increase is much less dramatic than for NGC 1600 M. This can be attributed to the much flatter original profile of this BCG. Once a core is present, it is much more difficult to scour and eject stars. 

The profiles for A2147-BCG M and NGC 6166 M are shown in Fig. \ref{fig:density_kicks_A2147} and Fig. \ref{fig:density_kicks_NGC6166} respectively. Unfortunately, due to their extremely long computational time, the kick simulations with $v_\mathrm{k}/v_\mathrm{e}$ of 0.9 for both galaxies are unfinished. However, in the case of the completed kicks, there is evidence of progressive core enlargement similar to NGC 1600 M. Interestingly, for the kick with $v_\mathrm{k}/v_\mathrm{e}=0.5$ in NGC 6166 M, a central cusp develops inside the flattened core as the SMBH reaches Brownian motion. We interpret this as the formation of a nuclear cluster around the SMBH. This phenomenon will be examined in detail in a separate paper (N. Khonji et al. 2024, in preparation).

Markov chain Monte Carlo (MCMC) fitting of the surface density profiles are shown in Tables \ref{tab:mcmc_scouring} and \ref{tab:mcmc_kicks}, for scouring and GW kicks, respectively. Using surface density as a proxy for luminosity, fitting was performed to the core-S\'ersic profile \citep{graham2003new}:
\begin{equation}
\label{eq:core-Sersic}
I = I' \Bigg[1+\bigg(\frac{r_\mathrm{b}}{r}\bigg)^\alpha\Bigg]^{\frac{\gamma}{\alpha}} \mathrm{exp} \,\Bigg(-b_n\bigg[\frac{r^\alpha + r_\mathrm{b}^\alpha}{r_\mathrm{e}^\alpha}\bigg]^{1/\alpha n}\Bigg) \ ,
\end{equation}
where $I$ is the luminosity, $\gamma$ is the inner logarithmic slope, $n$ is the S\'ersic index and $\alpha$ controls the sharpness of the transition between the inner power law and outer S\'ersic profiles. Here, $r_\mathrm{e}$ is the half-light radius of the profile outside the transition region, and $I'$ is related to $I_\mathrm{b}$, the intensity at $r_\mathrm{b}$, by:
\begin{equation}
\label{eq:idash}
I'=I_\mathrm{b} \, 2^{-\gamma/\alpha}\, \mathrm{exp}\,\Bigg[b_n\bigg(\frac{2^{1/\alpha}r_\mathrm{b}}{r_\mathrm{e}}\bigg)^{1 / n}\Bigg] \ .
\end{equation}

After the binary scouring phase (Table \ref{tab:mcmc_scouring}), the simulation of NGC 1600 achieved a core size of $0.68\kpc$, almost equal to the observational fit of $0.65\kpc$ (\citealt{dullo2019most}, Table \ref{tab:observed}). The corner plot for this fit is shown in Fig. \ref{fig:mcmc_ngc1600}, together with an overplot of the surface density profile using the fitted parameters, which shows a very good fit to the simulation data. For A2147-BCG, the simulated size of $1.32\kpc$ was also very close to that observed ($1.28\kpc$). However, for the larger core galaxies, NGC 6166 or A2261-BCG, the simulated cores sizes after scouring were considerably smaller than the observed ones, suggesting that an additional process is required to explain their core size. We note that all galaxies retain significant cuspiness, with slopes in the range 0.20 - 0.30. 

After the GW kicks (Table \ref{tab:mcmc_kicks}), the core sizes increase in all cases, and increase with increasing kick velocity for the galaxies with varying kick data available. Furthermore, $\gamma$ values reduce with increasing kick velocity for all galaxies as shown in Fig. \ref{fig:gamma_vk}. The slopes are also reduced in comparison to those after scouring. For the $v_\mathrm{k}/v_\mathrm{e}=0.5$ kick in NGC  6166 M, the central cusp, consistent with a nuclear cluster, was excluded from the fitting. A2261-BCG M achieves a size of $2.83\kpc$ after a kick of $v_\mathrm{k}/v_\mathrm{e}=0.5$, which is slightly larger than the observed size of $2.71\kpc$. The corner plot and overplot for this are shown in Fig. \ref{fig:mcmc_a2261}, with good fit to the data. The core size reaches a value of $3.62\kpc$ for a kick with $v_\mathrm{k}/v_\mathrm{e}=0.7$, but NGC 1600 M shows an even greater increase to $3.92\kpc$ for this kick velocity, increasing to an enormous $9.21\kpc$ for $v_\mathrm{k}/v_\mathrm{e}=0.9$. 

Fig. \ref{fig:rb_mbh} shows the relation between core size and black hole mass for the simulation results, in comparison to observational fits from \cite{dullo2019most}, \cite{dullo2014depleted} and \cite{rusli2013depleted}. The simulation core sizes are consistent with the observational data and most simulated galaxies achieve a core size similar to the observational one with scouring alone. Interestingly, however, NGC 6166 M and A2261-BCG M require a GW kick in order to obtain a core size as large as the observed one.
Furthermore, while our simulations show clearly that GW kicks produce flatter density profiles than scouring, we do not observe any clear relation between core size or black hole mass and the central slope $\gamma$ in the observed samples. These results are consistent with the expectation that GW kicks, though common, should be generally modest, and configurations leading to large recoil speeds should be rare. In the non-spinning case, the maximum GW recoil speed is $< 175 \kms$ for all $q$ \citep{gonzalez2007maximum}. Even with spin magnitudes as high as 0.8, aligned and anti-aligned with the orbital angular momentum, \cite{schnittman2007distribution} found recoil speeds would be $< 400 \kms$ for $q < 4$.

In order to investigate the physical reason for the flattening of the density profiles after GW recoil, we analyse the evolution of the stellar velocity anisotropy and populations in energy and angular momentum space. The velocity anisotropy parameter $\beta$ is given by:
\begin{equation}
    \label{eq:beta}
    \beta = 1 - \frac{\sigma_\theta^2 + \sigma_\phi^2}{2 \sigma_\mathrm{r}^2} ,
\end{equation}
where $\sigma_\theta$, $\sigma_\phi$ and $\sigma_\mathrm{r}$ are the components of the stellar velocity dispersion in spherical coordinates. Hence, $\beta = 0$ indicates isotropy, $\beta = 1$ that all orbits are radial, and $\beta = - \infty$ that they are circular. In Fig. \ref{fig:beta}, $\beta$ is plotted as a function of radius at three timepoints: the start of the merger (ICs), the end of the scouring phase ($t_\mathrm{k}$), and where the SMBH has reached Brownian motion (Br) at the end of the kick (or the most recent pericentre where the simulation is unfinished). This confirms that scouring does preferentially remove stars on radial orbits, but the effect of GW kicks appears dependent on their strength: kicks of $v_\mathrm{k}/v_\mathrm{e}=0.5$ appear to have the same degree of anisotropy or less than after scouring but, in general, kicks with $v_\mathrm{k}/v_\mathrm{e}$ of $0.7$ or $0.9$ preferentially remove additional stars on radial orbits centrally, increasing the anisotropy.

The total energy $E$ of stellar particles in NGC 1600 M is plotted against their angular momentum $J$ in Fig. \ref{fig:ej_scourkick} from the start of the simulation to the time of application of the GW kick (top row), and at various times following the GW kick of $v_\mathrm{k}/v_\mathrm{e}=0.9$ (bottom row). A significant change in the $E-J$ plane can be seen between the start of the simulation and time $t_f$ in the initial phase of the merger. By this time, the most central particles have the lowest energy and angular momentum. The latter is consistent with results from \citet{lagos2018quantifying}. Some of the low energy central particles are no longer present after hardening, a sign that close interactions with the BHB have ejected stars to larger radii through the slingshot process. This process slows down around the time the binary becomes hard, as most stars initially on low angular momentum orbits have been ejected, and we see little change from time $t_\mathrm{h}$ and the application of the GW kick. In this phase, only stars that are scattered to lower angular momentum orbits, typically from larger distances, interact with the binary. A further loss of low energy central particles can be seen in the bottom row of Fig.\ref{fig:ej_scourkick} after the GW recoil kick. In this phase, the SMBH makes repeated excursions to large radii and then back to the centre, displacing stars and causing a further reduction in central density. Typically, low angular momentum stars are removed from the core of the galaxy over time.

Particle tracking shows that higher energy particles have been ejected to larger radii during the initial merger, consistent with the expected increase in the effective radius. During hardening of the binary, there is outward movement of particles at most energies. This is even more pronounced during the GW kick, with central and low angular momentum particles moving to higher radii and gaining angular momentum. 

\begin{deluxetable*}{lcccccc}[htb!]
\tablecaption{MCMC core-S\'ersic fits after binary scouring.}
\label{tab:mcmc_scouring}
\tablehead{
    \colhead{Remnant} & \colhead{$r_\mathrm{b}$} & \colhead{$\gamma$} & \colhead{$\alpha$} & \colhead{n} & \colhead{$r_\mathrm{e}$} & \colhead{log $(\Sigma_\mathrm{b}/10^9 $M$_\odot $kpc$^{-2})$} \\
            & \colhead{[kpc]} &  &  &  & \colhead{[kpc]} & }
\startdata
NGC 1600 M  & $0.68\,^{+0.04}_{-0.05}$    & $0.30\,^{+0.04}_{-0.04}$    & $3.0\,^{+0.5}_{-0.4}$       & $6.0\,^{+0.1}_{-0.2}$       & $22.3\,^{+0.4}_{-0.7}$  & $23.07\,^{+0.05}_{-0.05}$           \\
A2147-BCG M & $1.32\,^{+0.15}_{-0.16}$    & $0.23\,^{+0.06}_{-0.07}$    & $1.4\,^{+0.2}_{-0.1}$       & $6.3\,^{+0.1}_{-0.1}$       & $24.7\,^{+2.5}_{-1.4}$  & $21.95\,^{+0.10}_{-0.09}$           \\ 
NGC 6166 M  & $0.90\,^{+0.09}_{-0.09}$    & $0.20\,^{+0.05}_{-0.06}$    & $1.15\,^{+0.05}_{-0.05}$    & $9.0\,^{+0.1}_{-0.1}$       & $60.1\,^{+0.2}_{-0.1}$  & $22.79\,^{+0.08}_{-0.08}$           \\
A2261-BCG M & $1.32\,^{+0.04}_{-0.05}$    & $0.24\,^{+0.02}_{-0.02}$    & $62.4\,^{+25.8}_{-28.7}$    & $2.15\,^{+0.02}_{-0.02}$    & $13.1\,^{+0.1}_{-0.1}$  & $22.98\,^{+0.02}_{-0.02}$           \\
\enddata
\end{deluxetable*}

\begin{deluxetable*}{lcccccccc}
\tablecaption{MCMC Core-S\'ersic fits after GW kicks.}
\label{tab:mcmc_kicks}    
\tablehead{
\colhead{Galaxy} & \colhead{$v_\mathrm{k}/v_\mathrm{e}$} & \colhead{$v_\mathrm{k}$} & \colhead{$r_\mathrm{b}$} & \colhead{$\gamma$} & \colhead{$\alpha$} & \colhead{n} & \colhead{$r_\mathrm{e}$} & \colhead{log $(\Sigma_\mathrm{b}/10^9 $M$_\odot $kpc$^{-2})$} \\
            & & \colhead{[km s$^{-1}$]} & \colhead{[kpc]} & & & & \colhead{[kpc]} &}
\startdata
NGC 1600 M  & $0.5$                         & 2121              & $2.48\,^{+0.09}_{-0.10}$  & $0.16\,^{+0.05}_{-0.05}$  & $5.4\,^{+0.8}_{-0.7}$ & $6.2\,^{+0.1}_{-0.1}$     & $22.5\,^{+0.2}_{-0.1}$    & $21.71\,^{+0.03}_{-0.03}$     \\
            & $0.7$                         & 2969              & $3.92\,^{+0.12}_{-0.13}$  & $0.16\,^{+0.04}_{-0.04}$  & $4.7\,^{+0.7}_{-0.6}$ & $6.1\,^{+0.1}_{-0.1}$     & $22.6\,^{+0.2}_{-0.2}$    & $21.11\,^{+0.03}_{-0.03}$     \\
            & $0.9$                         & 3818              & $9.21\,^{+0.89}_{-0.61}$  & $0.07\,^{+0.05}_{-0.05}$  & $2.5\,^{+0.7}_{-0.5}$ & $6.1\,^{+0.1}_{-0.1}$     & $22.5\,^{+0.2}_{-0.2}$    & $20.02\,^{+0.06}_{-0.07}$     \\ 
\noalign{\smallskip}
A2147-BCG M  & $0.5$                         & 1619              & $3.89\,^{+0.11}_{-0.11}$  & $0.16\,^{+0.03}_{-0.03}$  & $3.8\,^{+0.5}_{-0.4}$ & $6.3\,^{+0.1}_{-0.1}$     & $26.3\,^{+2.5}_{-2.4}$    & $20.81\,^{+0.02}_{-0.02}$     \\
            & $0.7$                         & 2267              & $5.15\,^{+0.13}_{-0.13}$    & $0.07\,^{+0.03}_{-0.03}$    & $3.6\,^{+0.4}_{-0.4}$ & $6.3\,^{+0.1}_{-0.1}$       & $25.8\,^{+2.7}_{-2.1}$    & $20.43\,^{+0.02}_{-0.02}$   \\
\noalign{\smallskip}
NGC 6166 M  & $0.5$                         & 2284              & $2.98\,^{+0.05}_{-0.04}$  & $0.02\,^{+0.03}_{-0.02}$   & $6.9\,^{+0.6}_{-0.5}$ & $9.07\,^{+0.02}_{-0.05}$     & $60.5\,^{+0.8}_{-0.4}$    & $21.09\,^{+0.01}_{-0.01}$ \\
            & $0.7$                         & 3198              & $4.86\,^{+0.10}_{-0.10}$  & $0.05\,^{+0.02}_{-0.02}$  & $3.2\,^{+0.3}_{-0.2}$ & $9.0\,^{+0.1}_{-0.1}$     & $68.7\,^{+7.4}_{-6.1}$    & $20.98\,^{+0.02}_{-0.02}$ \\
\noalign{\smallskip}
A2261-BCG M & $0.5$                         & 1772              & $2.83\,^{+0.09}_{-0.09}$  & $0.13\,^{+0.02}_{-0.02}$  & $5.4\,^{+1.0}_{-0.8}$ & $1.94\,^{+0.06}_{-0.03}$  & $13.1\,^{+0.1}_{-0.1}$    & $22.32\,^{+0.02}_{-0.02}$     \\
            & $0.7$                         & 2498              & $3.62\,^{+0.12}_{-0.13}$  & $0.12\,^{+0.02}_{-0.02}$  & $3.7\,^{+0.4}_{-0.3}$ & $2.1\,^{+0.1}_{-0.1}$     & $13.0\,^{+0.1}_{-0.1}$    & $22.02\,^{+0.02}_{-0.02}$     \\
            & $0.9$                         & 3212              & $5.32\,^{+0.17}_{-0.17}$  & $0.10\,^{+0.03}_{-0.03}$  & $3.3\,^{+0.4}_{-0.4}$ & $2.05\,^{+0.10}_{-0.10}$  & $13.0\,^{+0.1}_{-0.1}$    & $21.01\,^{+0.02}_{-0.02}$     \\
\enddata
\end{deluxetable*}

\section{Discussion and Conclusions}

We have studied the physical processes of black hole binary scouring and gravitational wave (GW) recoil, which together may result in the formation of very large cores, as observed in a few of the most massive elliptical galaxies.

We find that binary scouring alone can form large cores of up to $\sim1.3\kpc$, such as those observed in NGC 1600 and A2147-BCG. However, an additional process is required to form the largest cores, greater than $2\kpc$, observed in galaxies such as NGC 6166 and A2261-BCG. We have shown that GW recoil can form these very large cores with  kicks of less than half of the escape speed $v_\mathrm{e}$ for the $2.11\kpc$ core of NGC 6166 and the $2.71\kpc$ core of A2261-BCG. This corresponds to kick velocities of less than $\sim 2300\kms$  and $\sim 1800\kms$, respectively, well below the theoretical maximum recoil for spin-dominated kicks from numerical relativity simulations, found by \cite{campanelli2007maximum} to be $\sim 4000\kms$ and by \cite{lousto2019kicking} to be $\sim 5000\kms$. Indeed, the latter found the probability of a recoil kick in the range $1000 - 2000\kms$ to be $\sim$ 5\%. We note that, in the case of such spin-dominated kicks, equal-mass mergers have higher recoil velocities than smaller mass ratio mergers. If we define the mass ratio as $q \leq 1$, the expected scaling is $v_\mathrm{k} \propto q^2$ \citep{campanelli2007large}. 

We also find that GW recoil leads to flatter inner density profiles than binary scouring alone. This occurs rapidly after the kick and is apparent by the first apocentre. In general, the values of the inner slope
$\gamma$ found after GW kicks are much closer to those of the observed profiles than 
those with binary scouring alone. This might imply that all the studied galaxies have experienced some GW recoil in addition to scouring. However, the degeneracies involved in fitting a six-parameter model such as the core-S\'ersic profile with an MCMC procedure make this hard to claim definitively.

\cite{nasim2021formation} performed multiple simulations of core formation in A2261-BCG. However, only one used the same parameters including $M_\bullet$ used in this study. In that simulation, they found a core radius of $0.9\kpc$ at the end of scouring, $2.64\kpc$ after a GW kick with $v_\mathrm{k}/v_\mathrm{e} = 0.3$, and $2.96\kpc$ after a $v_\mathrm{k}/v_\mathrm{e} = 0.8$ kick. We found a slightly larger core after scouring ($1.32\kpc$), a core size between their $v_\mathrm{k}/v_\mathrm{e} = 0.3$ and $v_\mathrm{k}/v_\mathrm{e} = 0.8$ kicks for our $v_\mathrm{k}/v_\mathrm{e} = 0.5$ kick ($2.98 \kpc$), and a larger core after our $v_\mathrm{k}/v_\mathrm{e} = 0.7$ kick ($3.62\kpc$) than their $v_\mathrm{k}/v_\mathrm{e} = 0.8$ kick. The differences are likely due to differences in the fitting procedure.

Our results fit the $r_\mathrm{b} - M_\bullet$ relation of observed core galaxies well (Fig. \ref{fig:rb_mbh}). Since the GW recoil velocity scales with the masses of the SMBHs \citep{campanelli2007maximum}, one could expect galaxies with more massive SMBHs to display flatter cores. However, we find no clear relation between $M_\bullet$ and $\gamma$ in the observed data. This might be due to the rarity of large recoils, which require both large SMBH masses and particular spin configurations, or due to the fitting procedure being unable to robustly measure gamma over the region of interest. We will examine this in more detail in forthcoming work.

Finally, we find that high GW kicks preferentially remove low angular momentum stars from cores. Although this is already known to occur in binary scouring, kicks with $v_\mathrm{k}/v_\mathrm{e}$ of $0.7$ or higher can further remove these stars, leading even greater central anisotropy.

Overall, we find GW kicks are a very plausible cause for the formation of the largest cores. Since two of the modelled SMBHs are insufficient to form the observed core sizes by scouring alone, without an additional mechanism the SMBH masses would have to lie well above the $r_\mathrm{b} - M_\bullet$ relation, potentially $\gtrsim 10^{11} \msun$. GW recoil avoids invoking such hypermassive black holes. Indeed, core sizes greater than $\sim 2 \kpc$ may indicate that significant GW recoil has occurred, and flat cores may be a marker for some degree of recoil. 

However, other mechanisms of large core formation have been proposed: stalling infallers could form very large cores as additional nuclear components rather than by central deficits \citep{goerdt2010core, bonfini2016quest}; resonant interactions between SMBHs and multiple recoil events could form large cores in multiple SMBH systems \citep{kulkarni2012formation}. We will consider such processes in future work.  

\begin{acknowledgments}
NK is supported by a grant from the Science and Technology Facilities Council (STFC). AG would like to thank the STFC for support from grant ST/Y002385/1. JIR would like to acknowledge support from STFC grants ST/Y002865/1 and ST/Y002857/1.
\end{acknowledgments}

%




\appendix

\section{Density profiles for A2147-BCG and NGC 6166}

\begin{figure*}[htb!]
    \includegraphics[width = \textwidth]{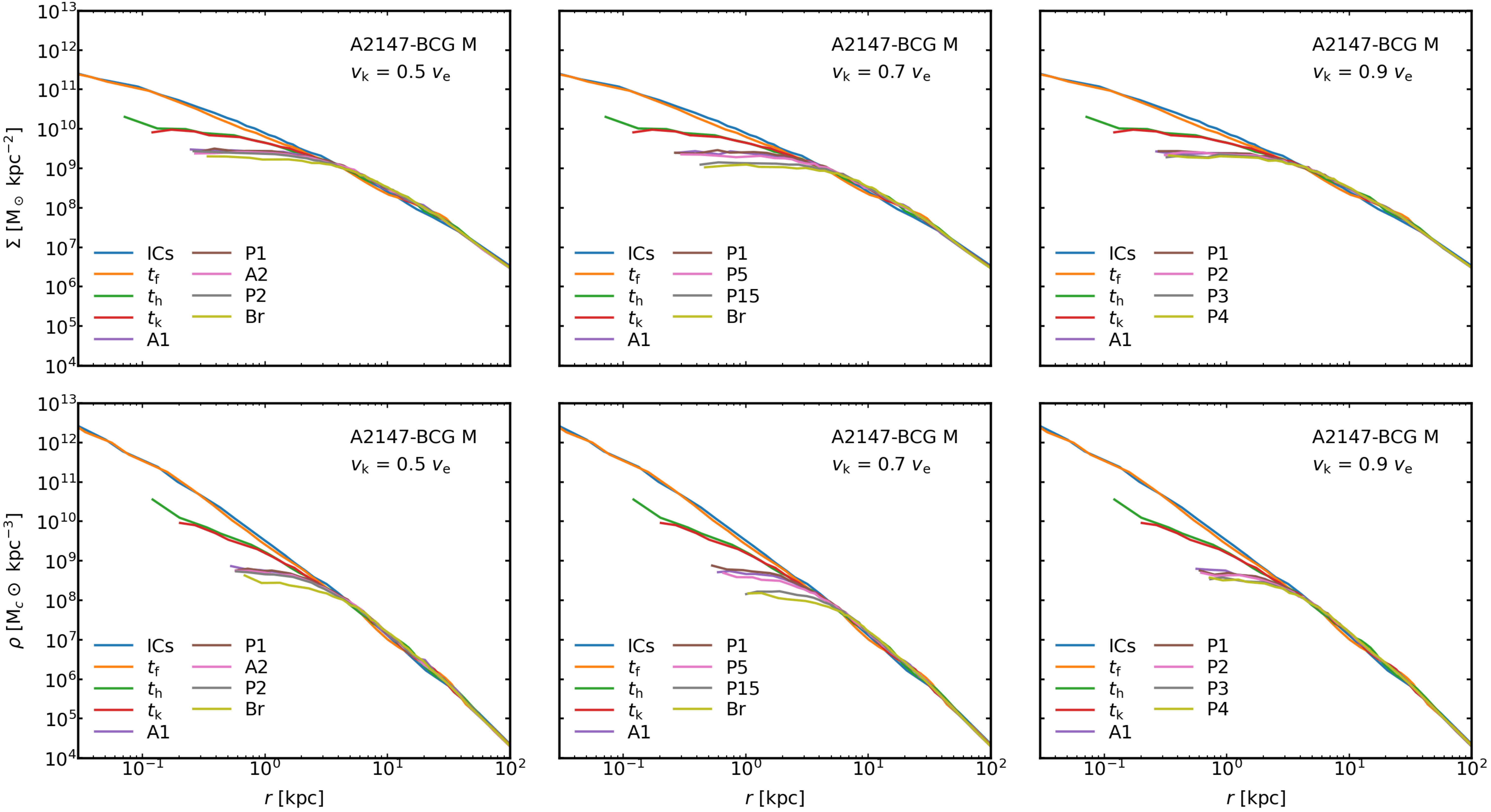}
     \caption{Surface density (top row) and spatial density (bottom row) profiles during merger and after GW kick for A2147-BCG. $v_\mathrm{k}/v_\mathrm{e}$ 0.5 (1st col.), 0.7 (2nd col.) and 0.9 (3rd col.). ICs indicates start of galaxy merger, $t_\mathrm{k}$ = 0 is time of SMBH merger. `A' and `P' indicate  apocentre and pericentre respectively, and are followed by the number of passages. `Br' indicates the SMBH remnant has settled into Brownian motion.}
    \label{fig:density_kicks_A2147}
\end{figure*}

\begin{figure*}
    \includegraphics[width = \textwidth]{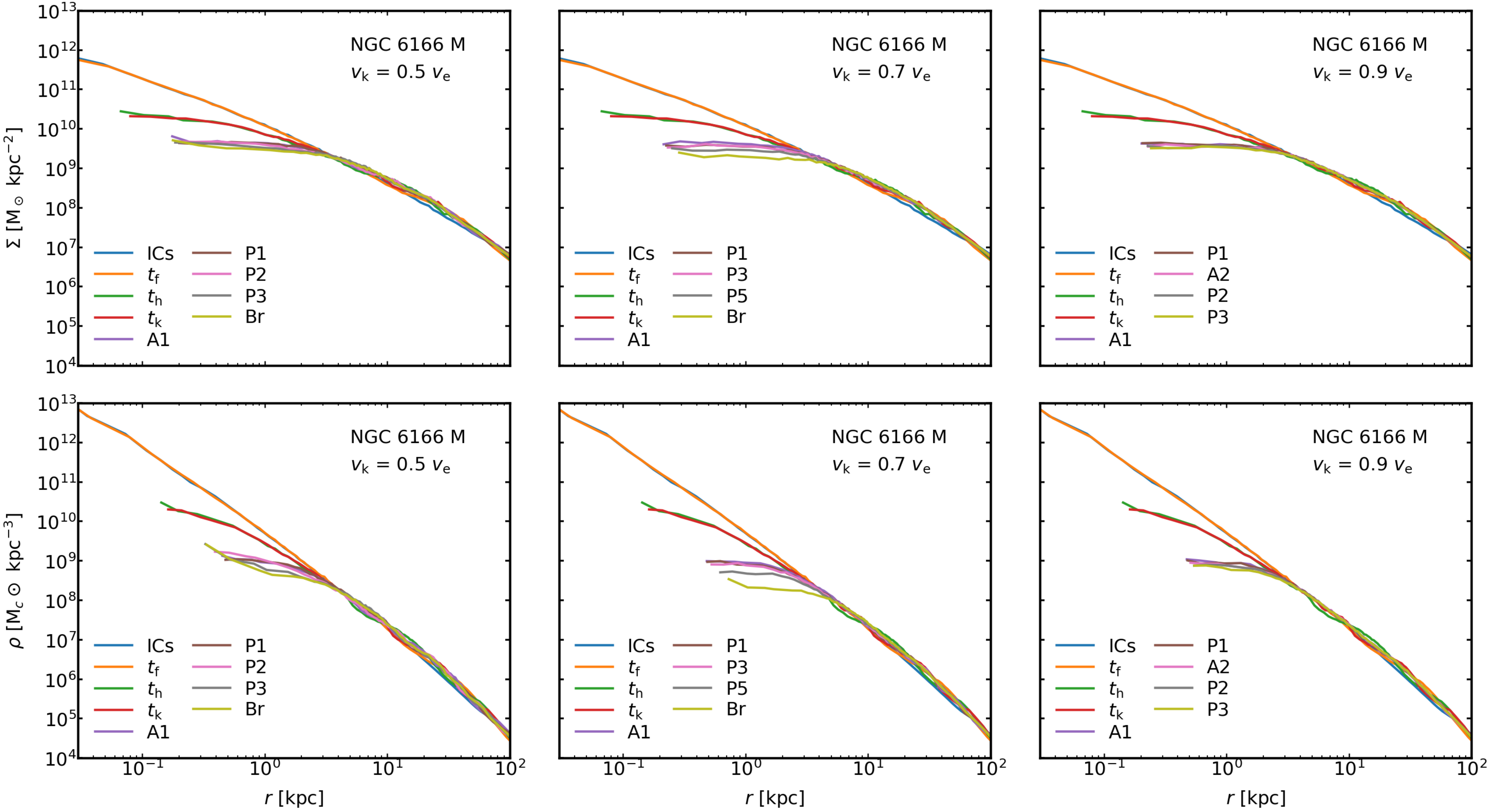}
     \caption{As for Fig. \ref{fig:density_kicks_A2147}, but for NGC 6166 M.}
    \label{fig:density_kicks_NGC6166}
\end{figure*}

Fig. \ref{fig:density_kicks_A2147} shows the surface density (top row) and volume density (bottom row) profiles for A2147-BCG M during both binary scouring and for the entirety of each GW kick. Fig. \ref{fig:density_kicks_NGC6166} shows the corresponding profiles for NGC 6166 M. 

Due to their long computational time, the $v_\mathrm{k}/v_\mathrm{e}$ of 0.9 runs are unfinished. However, where completed, there is progressive core enlargement similar to NGC 1600 M (Fig. \ref{fig:density_kicks_NGC1600}). Furthermore, for the $v_\mathrm{k}/v_\mathrm{e}$ of 0.3 runs, there is increased flattening at the first apocentre for both galaxies, and the second for A2147-BCG M, compared to later times. We attribute this to the higher bound mass at low $v_\mathrm{k}/v_\mathrm{e}$ removing stars from the centre in the initial phases, some of which is subsequently returned.  

\clearpage
\section{Corner plots for MCMC fits}

\begin{figure*}[htb!]
    \centering
        \includegraphics[width = \textwidth]{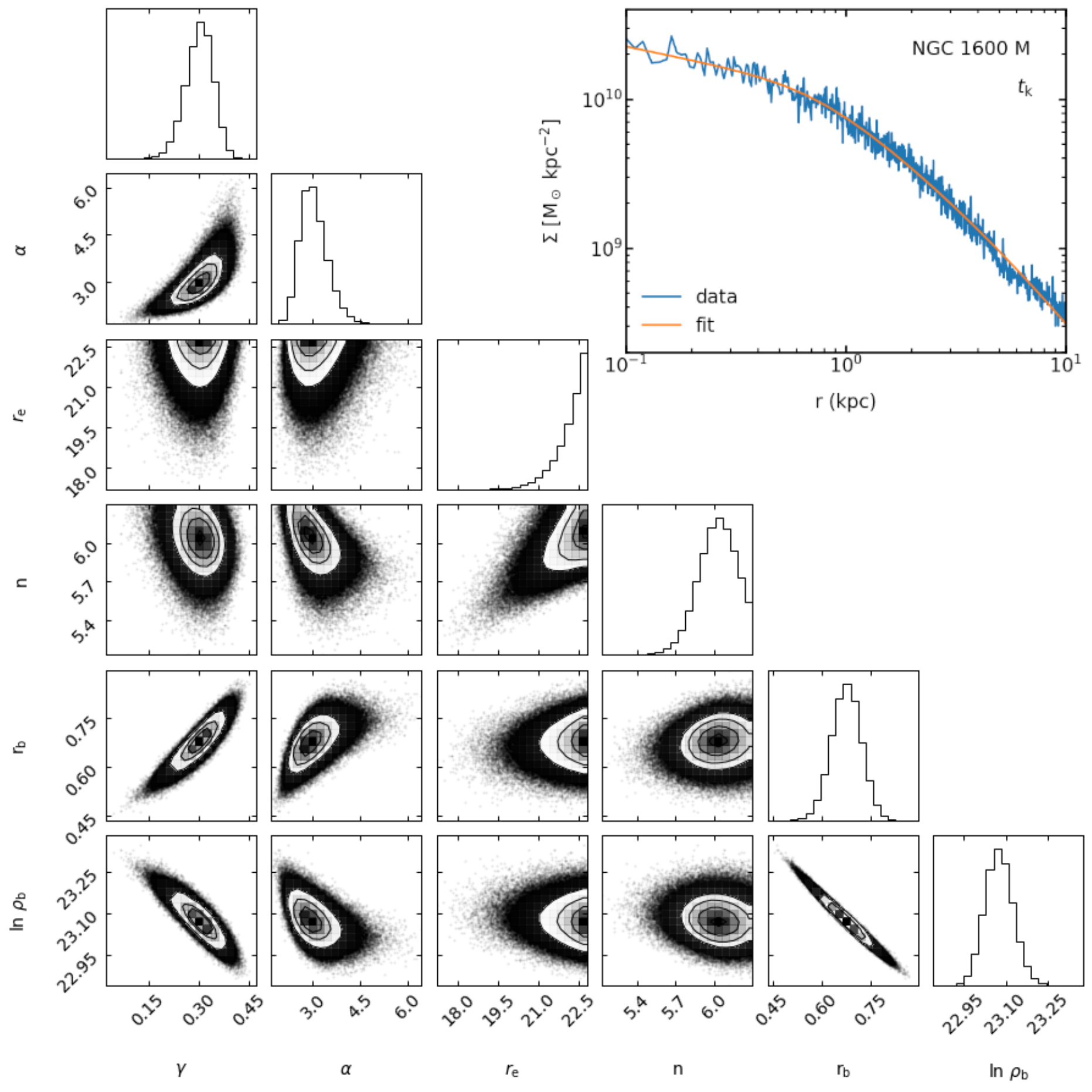}
    \caption{Corner plot resulting from the MCMC core-S\'ersic fit to NGC 1600 M after binary scouring. The inset panel shows the surface density profile obtained from the $N$-body data compared with the analytic fitted profile.}
    \label{fig:mcmc_ngc1600}
\end{figure*}

\begin{figure*}[htb!]
    \centering
        \includegraphics[width = \textwidth]{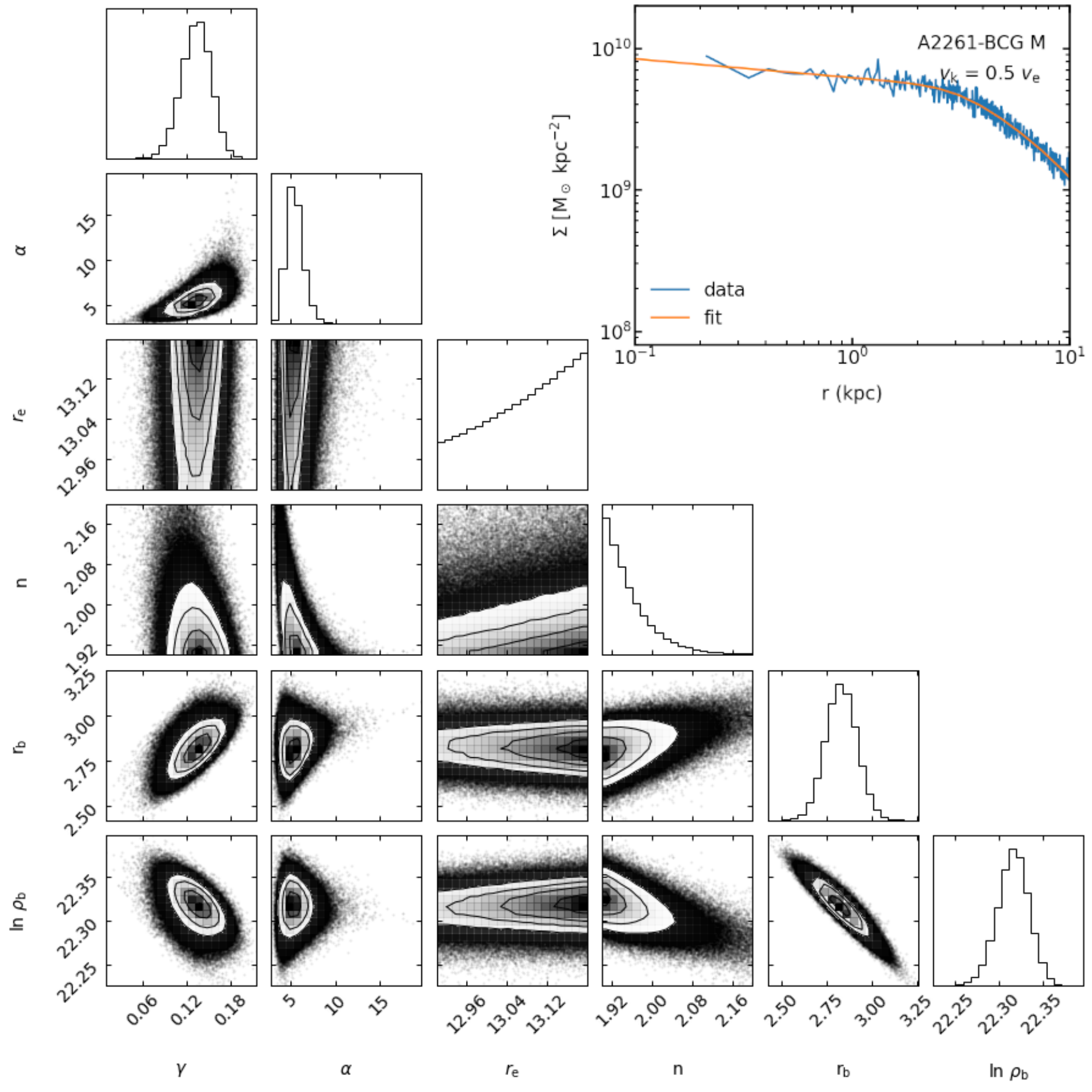}
    \caption{Corner plot resulting from the MCMC core-S\'ersic fit to A2261-BCG M after the GW kick $v_\mathrm{k}/v_\mathrm{e}=0.5$ is applied. The inset panel shows the surface density profile obtained from the $N$-body data compared with the analytic fitted profile.}    
    \label{fig:mcmc_a2261}
\end{figure*}

Fig. \ref{fig:mcmc_ngc1600} shows the corner plot for the MCMC core-S\'ersic fit to the surface density profile of the NGC 1600 M remnant after the binary scouring phase. The core size of $0.68\kpc$ is almost equal to the observational fit of $0.65\kpc$ (\citealt{dullo2019most}, Table \ref{tab:observed}). The inset shows an overplot of the surface density profile using the fitted parameters, which shows a very good fit to the simulation data.

Fig. \ref{fig:mcmc_a2261} shows the corner plot for the MCMC core-S\'ersic fit to the surface density profile of A2261-BCG M remnant after a kick of $v_\mathrm{k}/v_\mathrm{e}=0.5$. The core achieves a size of $2.83\kpc$, which is slightly larger than the observed size of $2.71\kpc$. The overplot again fits the data well.

\clearpage
\bibliography{core_formation}{}
\bibliographystyle{aasjournal}



\end{document}